\documentclass{jfm}
\usepackage{graphicx}
\usepackage{epstopdf, epsfig}
\usepackage{subfigure}
\usepackage{epsf}
\usepackage{bm}
\usepackage{multirow}
\usepackage{float}
\usepackage{xcolor}

\shorttitle{Dynamics of zigzagging/spiralling bubble pairs}
\shortauthor{J. Zhang, M.-J. Ni, J. Magnaudet}

\title{Three-dimensional dynamics of a pair of deformable bubbles rising initially in line. Part 2: Highly inertial regimes}

\author{Jie Zhang\aff{1}, 
Ming-Jiu Ni\aff{1,2}
\corresp{\email{mjni@ucas.ac.cn}}
 \and Jacques Magnaudet\aff{3}
\corresp{\email{Jacques.Magnaudet@imft.fr}}}

\affiliation{\aff{1}State Key Laboratory for Strength and Vibration of Mechanical Structures, School of Aerospace, Xi'an Jiaotong University, Xi'an, China
\aff{2}School of Engineering, University of Chinese Academy of Sciences, Beijing, China
\aff{3}Institut de M\'ecanique des Fluides de Toulouse (IMFT), Universit\'e de Toulouse, CNRS, Toulouse, France}

\begin{document}

\maketitle

\begin{abstract}
The buoyancy-driven dynamics of a pair of gas bubbles released in line is investigated numerically,  
focusing on highly inertial conditions under which isolated bubbles follow non-straight paths. In an early stage, the second bubble always drifts out of the wake of the leading one. Then, depending on the ratios of the buoyancy, viscous and capillary forces which define the Galilei ($Ga$) and Bond ($Bo$) numbers of the system, five distinct regimes specific to such conditions are identified, in which the two bubbles may rise independently or continue to interact and possibly collide in the end. In the former case, they usually perform large-amplitude planar zigzags within the same plane or within two distinct planes, depending on the oblateness of the leading bubble. However, for large enough $Ga$ and low enough $Bo$, they follow nearly vertical paths with small-amplitude erratic horizontal deviations. Increasing $Bo$ makes the wake-induced attraction toward the leading bubble stronger, forcing the two bubbles to realign vertically one or more times along their ascent. During such sequences, wake vortices may hit the trailing bubble, deflecting its path and, depending on the case, promoting or hindering further possibilities of interaction. In some regimes, varying the initial distance separating the two bubbles modifies their lateral separation beyond the initial stage. Similarly, minute initial angular deviations favour the selection of a single vertical plane of rise common to both bubbles. These changes may dramatically affect the fate of the tandem as, depending on $Bo$, they promote or prevent future vertical realignments.



\end{abstract}

\section{Introduction}\label{sec1}

In the first part of this investigation (Zhang, Ni \& Magnaudet (2021), hereinafter referred to as ZNM), we analyzed the results of a series of simulations revealing the mechanisms governing the hydrodynamic interactions between two deforming gas bubbles released in line in a liquid at rest. The physical parameters were selected in such a way that the bubbles rose at moderate Reynolds number, and each bubble taken separately would ascend in a straight line. However, millimeter-size air bubbles rising in low-viscosity liquids, most notably in water, are subject to path instability. Consequently, they usually follow either zigzagging planar paths or more or less flattened spiralling paths, with in both cases large-amplitude horizontal excursions. These are the regimes on which this second part focuses.\\
\indent As discussed in ZNM, interactions between two neighbouring bubbles (so-called pair interactions) play a key role in the microstructure of bubbly suspensions, as they govern to a large extent the bubble distribution and the agitation in the carrying liquid. Early computational studies of buoyancy-driven bubbly suspensions \citep{Smereka1993, Sangani1993} assumed the bubbles to keep a spherical shape and disregarded any possible influence of viscosity. Since the potential flow approximation predicts that two bubbles rising in line repel each other while they attract each other when rising side by side, such simulations inescapably concluded that the suspension dynamics lead to the formation of large horizontal bubble clusters. The next generation of simulations addressed more realistic conditions by considering the full Navier-Stokes equations, possibly including surface tension effects.
 With Reynolds numbers of $\mathcal{O}(10)$ to $\mathcal{O}(100)$ and spherical or weakly deformed bubbles, these simulations confirmed the tendency of bubble pairs to align horizontally, albeit less clearly than in the potential flow approximation \citep{Esmaeeli1999, bunner2002dynamics,esmaeeli2005direct,yin2008velocity}. With significantly deformed bubbles, the dynamics of bubbly suspensions was observed to depend crucially on the Reynolds number. More specifically, the simulations of \cite{esmaeeli2005direct} with Reynolds numbers of $\mathcal{O}(100)$ revealed quite homogeneous bubble distributions, while for Reynolds numbers of $\mathcal{O}(10)$, marked vertical bubble alignments were noticed by \cite{bunner2003effect}, suggesting that some `chimney effect' is at work under such conditions. It must be pointed out that, for technical reasons, all of the above studies made use of some simplifying assumptions which depart from common physical conditions. In most of them, the liquid-to-gas density ratio was reduced by a factor of $20$ to $50$, making the bubble motion less sensitive to small fluctuations of the carrying liquid than under standard experimental conditions, while in others \citep{yin2008velocity} bubbles were not allow to deform. These restrictions were removed in the most recent simulations \citep{loisy2017buoyancy,innocenti2021direct}. However, the parameters were still chosen in such a way that bubbles are not subject to path instability. That is, the liquid viscosity and/or surface tension were sufficiently large to prevent an isolated bubble from rising along a zigzagging or spiralling path, making the predictions barely representative of most experiments performed in water. Even with the largest inertia-to-viscosity force ratio considered to date \citep{innocenti2021direct}, surface tension was selected so as to keep the bubbles only mildly deformed, preventing the occurrence of path instability. About one billion grid points were employed to track the flow generated by the rise of 256 bubbles in that study, but a ten times larger grid would be required to deal with regimes involving zigzagging or spiralling bubbles, owing to the very thin boundary layers and complicated wake structures involved. \\
\indent Despite these demanding requirements, detailed numerical investigations of this highly inertial regime are needed, in view of its specificities and its ubiquity in natural and engineering air-water bubbly flows. For instance, it is known that the liquid velocity fluctuations generated by the rise of a dilute bubble swarm are highly anisotropic when the bubbles are only slightly distorted, the vertical fluctuations having a variance $4-5$ times larger than their horizontal counterpart \citep{Zenit2001}. In contrast, this ratio falls to values close to $2$ for larger bubbles exhibiting path instability \citep{risso2002velocity,Riboux2010}. Obviously, the reason for this sharp decrease stands in the large horizontal fluctuations of bubble positions along a zigzagging or spiralling path, which in turn induce large horizontal fluctuations in the liquid velocity. More importantly, the fraction of the liquid agitation resulting from the interacting bubble wakes is much larger in the case of zigzagging/spiralling bubbles. This contribution makes the whole carrying flow exhibit genuine turbulent properties over a significant range of scales. The mixing ability of this `bubble-induced turbulence' is well highlighted by examining how a weakly diffusive species (e.g. dye) or a temperature gradient is mixed in a bubble swarm \citep{Almeras2015,Gvozdic2018}. In both cases, a clear increase of the relevant transfer coefficient with the gas volume fraction is observed in very dilute swarms (i.e. gas volume fractions of the order of $1\%$), together with a dramatic increase in the strength of the scalar fluctuations.


Although still far from such turbulent bubbly suspensions and the key statistical information gained from the aforementioned studies, the simplest arrangement capable of providing detailed insight into wake interactions in the relevant highly inertial regime is that involving a pair of bubbles. Focusing on such an elementary configuration offers a complementary point of view with respect to a many-bubble configuration, since the hydrodynamic mechanisms involved in the interaction process can be identified and analyzed in a deterministic manner, and the few control parameters can be varied separately to check their influence. We refer the reader to the introduction of ZNM for a review of the available knowledge regarding the dynamics of pairs of spherical or weakly distorted bubbles in moderately inertial regimes as well as in the potential flow approximation. In the regime of interest here, some experimental studies considered the side-by-side arrangement \citep{Duineveld1998,Sanada2009,kong2019hydrodynamic}. In this configuration, the potential flow approximation predicts that the two bubbles are attracted toward each other. However, when they get close enough, their wakes interact directly and the sign of the transverse force acting on each bubble may reverse. This is why one objective of these studies was to check the predictions of the inviscid theory of \cite{Chesters1982} regarding the conditions under which the two bubbles bounce (possibly repeatedly) or rather coalesce during their ascent. It must be noticed that none of these studies addressed quantitatively the detail of wake interactions. Only one of them  \citep{Sanada2009} visualized the wakes through a photochromic technique and could relate the reversal of the transverse bubble motion to the encounter of the two wakes. \\
\indent Some experiments were also carried out in the in-line configuration on which the present investigation focuses, the most recent one being that of \cite{kusuno2019lift}. However, all these studies considered Reynolds numbers of $\mathcal{O}(10-100)$ which belong to the moderately inertial regime examined numerically by ZNM and, independently but with the same code, by \cite{kusuno2021wake}. A noticeable exception is the work of \cite{filella2020interaction} in which wake interactions behind two bubbles rising in a thin-gap cell were scrutinized using time-resolved particle image velocimetry. However the corresponding two-dimensional dynamics make the problem barely comparable to the three-dimensional configuration of interest here. The main outcome of the recent studies by \cite{kusuno2019lift}, \cite{kusuno2021wake} and ZNM is the existence of three regimes with markedly different dynamics, according to the values of the control parameters and to the detail of the initial conditions. In short, pairs of nearly-spherical bubbles released exactly in line and rising with a Reynolds number of $\mathcal{O}(10)$ follow the Drafting-Kissing-Tumbling (DKT) scenario widely observed with sedimenting spherical particles \citep{joseph1986nonlinear, fortes1987nonlinear}. Conversely, pairs of bubbles exhibiting a sufficient oblateness always collide and eventually coalesce, most of the time in the head-on configuration. Below this critical, Reynolds-number dependent oblateness and beyond the narrow parameter range where the DKT mechanism takes place, the two bubbles follow an Asymmetric Side Escape (ASE) scenario in which the trailing bubble leaves the wake of the leading bubble through a vigorous lateral drift. Beyond this crucial stage, the trailing bubble rises along a new nearly vertical path, whereas the leading bubble is barely disturbed by the interaction and essentially goes on rising along its initial path, with however some marginal inclination. The tandem stabilizes in a configuration whose final geometry, i.e. lateral separation and inclination of the line of centres with respect to the vertical, depends significantly on the control parameters and is very sensitive to initial conditions. Since new bubble pairs form continuously in a swarm, the DKT and ASE scenarios are self-repeating in a real suspension. Given that the latter is by far the most frequent in the moderately inertial regime and yields significantly different bubble pair geometries according to tiny changes in the initial conditions, it is no surprise that bubbly suspensions with the corresponding characteristics are far more homogeneous than predicted by the simplistic potential flow approximation.\\
\indent Still considering the in-line configuration, our aim in this second part is to explore the regime in which bubbles rise with Reynolds numbers of $\mathcal{O}(100-1000)$. In most cases, an isolated bubble would then either perform large-amplitude planar zigzags or follow a spiralling path. 
Similar to ZNM, we vary the control parameters and the initial conditions so as to cover a significant range of physical conditions and analyze different evolution scenarios, with a specific attention on those involving the direct interaction between the trailing bubble and the three-dimensional time-dependent wake of the leading bubble.
We present the problem and summarize the numerical approach in \S\,\ref{sec2} (a specific test aimed at confirming that the grid resolution is adequate even for the highest Reynolds numbers considered is discussed in Appendix \ref{app:sec1}). Then the results obtained under nominal initial conditions are discussed in \S\,\ref{sec3}. Influence of the initial conditions, i.e. a slight angular deviation or a change in the vertical separation, is discussed in \S\,\ref{sec4}. A summary of the main findings and some prospects are presented in \S\,\ref{sec5}.

\section{Problem statement and outline of the numerical approach}\label{sec2}

Since the problem was introduced in detail in ZNM, only a brief account is presented here. A pair of initially spherical bubbles with radius $R$ is released in line with a vertical centre-to-centre separation $S_0$. Starting from rest, the two bubbles rise freely under the effect of buoyancy. The evolution of the tandem geometry is tracked by considering the vertical and horizontal dimensionless separations $\overline{S}(t)=S(t)/R$ and $\overline{S}_r(t) = S_r(t)/R$, respectively, and the angular deviation of the line of centres with respect to the vertical, $\theta(t)$ (see figure~1 in ZNM).
In addition to the initial separation $\overline{S}_0=S_0/R$ (hereinafter set to $\overline{S}_0 = 8$ except specified otherwise), the flow and bubble dynamics are characterized by the Galilei and Bond numbers respectively defined as
\begin{equation}
  Ga = \rho_l g^{1/2}R^{3/2}/\mu_l\,, \hspace{0.5cm} Bo = \rho_l gR^2/\gamma\,,
\end{equation}
where $\rho_l$ and $\mu_l$ are the density and viscosity of the carrying liquid, $\gamma$ is the surface tension, and $g$ denotes gravity. Once the terminal velocity $u_T$ of each bubble is known, the terminal Reynolds and Weber numbers, $Re = \rho_lu_TR/\mu_l$ and $We = \rho_lu_T^2R/\gamma$, may be determined. Moreover, the Morton number $Mo=Bo^3/Ga^4= g\mu_l^4/\rho_l\gamma^3$ is useful to identify the dynamics of bubble pairs in a given fluid, irrespective of the bubble size. In ZNM, $Ga$ and $Bo$ were varied in the range $10 \leq Ga \leq 30$ and $0.01 \leq Bo \leq 1.0$, respectively. The corresponding terminal Reynolds numbers were such that $10 \lesssim Re \lesssim 120$, so that an isolated bubble followed a rectilinear path. 
Here we consider the parameter range $30 < Ga \leq 90$ and $0.02 \leq Bo \leq 1.0$. In that range, an isolated bubble is likely to follow most of the time a zigzagging or a (possibly flattened) spiralling path according to the phase map of \cite{cano2016paths}. Under such conditions, the bubble wake is no longer axisymmetric. Rather, the wake is dominated by a pair of  counter-rotating trailing vortices in which the streamwise vorticity is concentrated. As significant part of the paper focuses on the role of these trailing vortices on the dynamics of the bubble pair, especially on the direct interaction in some regimes of the trailing bubble with the vortex pair emanating form the leading bubble.\\
\indent 
The results to be discussed below are obtained by solving the three-dimensional time-dependent Navier-Stokes equations valid throughout the flow (i.e. including capillary effects) with the open source flow solver \textit{Basilisk} (http://basilisk.fr) described in \cite{popinet2009accurate,popinet2015quadtree}. The numerical schemes employed in this solver are detailed in the above two references and are summarized in ZNM. In particular, a geometrical volume of fluid approach is employed to track and advance the liquid-gas interface. A sophisticated adaptive mesh refinement technique makes it possible to locally refine the grid close to the interface and within high vorticity regions, a feature that greatly enhances the computational efficiency. In ZNM, an additional grid refinement strategy was activated to capture the flow within the thin gap separating bubbles under near-contact conditions. Such conditions are met here for Bond numbers of $\mathcal{O}(1)$. However, since the focus of the present paper is on the role of the leading bubble wake in the dynamics of the tandem, collisions are not examined in detail. That is, the aforementioned thin-film grid refinement strategy is not activated, making us unable to distinguish between  the collisions that under real conditions are followed by a bounce of the two bubbles and those leading to their coalescence. \\
\indent Similar to ZNM, we make use of a cubic numerical domain, with size $(240R)^3$. This large size guarantees that artificial confinement effects are negligibly small throughout the considered range of parameters. A free-slip condition is imposed on all four lateral boundaries, while a periodic condition is prescribed on the top and bottom boundaries. 
The spatial resolution is refined down to $\Delta_{min} = R/68$ close to the bubble interface and to $\Delta = R/17$ in the wake. With this discretization, ZNM established that the evolutions of the rising speed of each bubble are grid-independent up to $Ga=30$. However, the highest Reynolds numbers considered here are typically four times larger than in ZNM. Therefore the boundary layers may be twice as thin, making it necessary to check whether or not the above resolution remains sufficient under such conditions. To this aim, a grid convergence test was carried out with the set of parameters $Ga=90,\,Bo=0.05$ for which the Reynolds number of an isolated bubble is close to $470$. The results of this test are discussed in Appendix \ref{app:sec1}, from which it can be concluded that the above resolution properly captures the details of the flow throughout the Reynolds number range considered here. \\
\indent In the course of the paper, we shall often refer to the study of \cite{cano2016paths} (hereinafter referred to as CL16) who computed the path of isolated bubbles close to the onset of path instability. These computations were carried out with the \textit{Gerris} open solver. This code is the direct predecessor of \textit{Basilisk} and makes use of the same algorithms. The grid resolution employed in CL16 was similar to the present one. These remarks are important because they imply that the results of this previous study can be safely used as a reference to compare the evolution of a bubble pair computed here with that of the corresponding isolated bubble reported in CL16.  Throughout the paper, all results are normalized using the characteristic length $R$ and time $\sqrt{R/g}$. Bubble deformation is characterized by the aspect ratio $\chi = b/a$, with $b$ and $a$ the length of the major and minor axes, respectively.


\section{Results in the reference case}\label{sec3}

\subsection{Overview}\label{sec3.1}

\begin{figure}
\vspace{5mm}
\centering
  \includegraphics[width=0.98\textwidth]{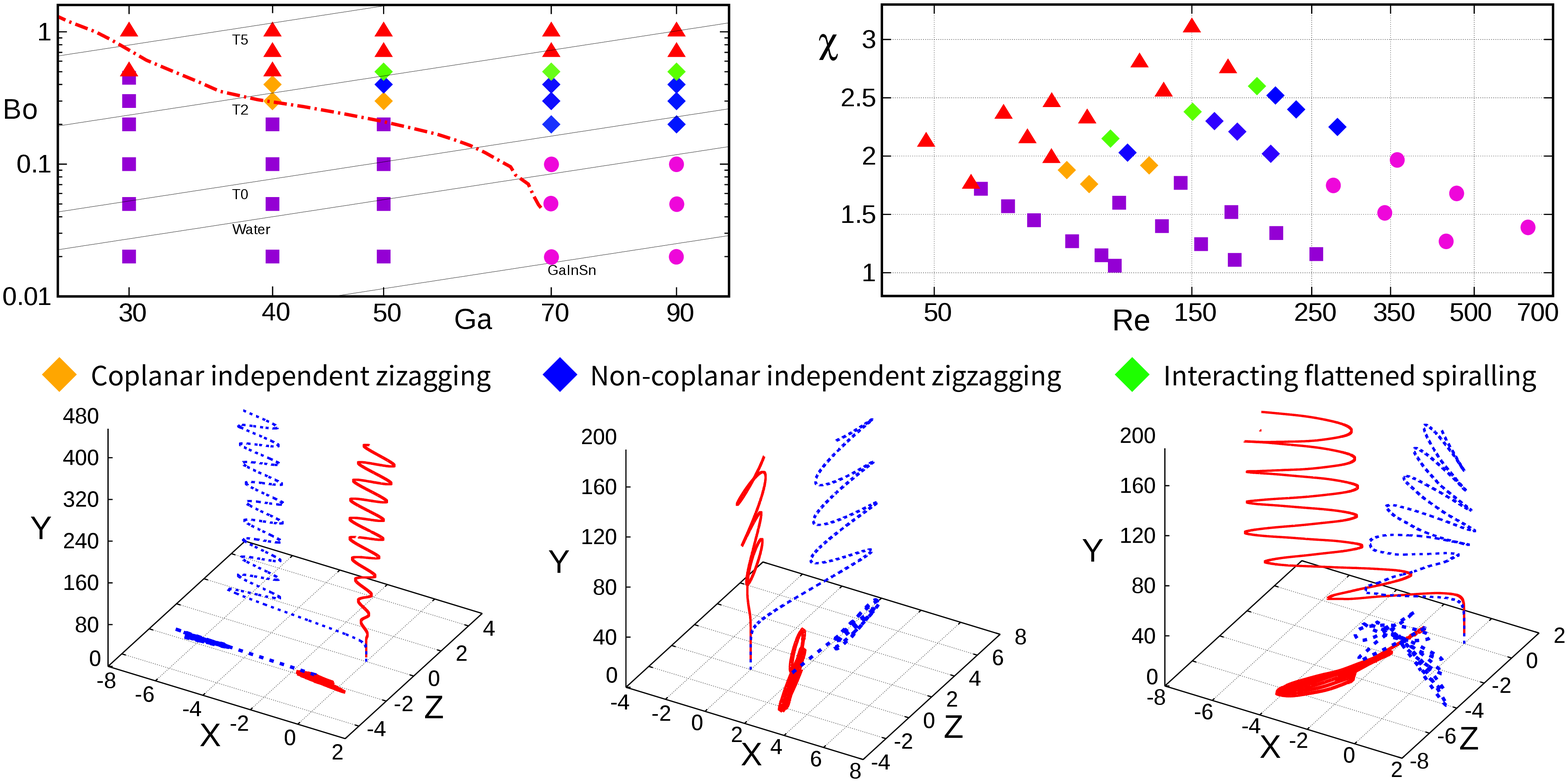}
\caption{Configuration map based on simulation results: $(a)$ $(Ga, Bo)$-plane; $(b)$ $(Re, \chi)$-plane. $\color{violet}{\blacksquare}$: asymmetric side escape (ASE) regime; $\color{magenta}{\bullet}$: small-amplitude independent erratic (SIE) regime; $\color{red}{\blacktriangle}$: collision regime; $\color{orange}{\blacklozenge}$, $\color{blue}{\blacklozenge}$, $\color{green}{\blacklozenge}$: CIZ, NIZ and IFS oscillating regimes, respectively. $(c)$: front and top views of the paths corresponding to the last three regimes for $(Ga, Bo) =$ (50, 0.3), (70, 0.3) and (90, 0.5), respectively (the red solid and blue dashed lines refer to the LB and TB, respectively). The dash-dotted line in $(a)$ is the neutral curve corresponding to the onset of path instability for an isolated bubble (CL16); the thin solid lines are the iso-$Mo$ lines corresponding to different liquids, with the GaInSn liquid metal (Galinstan) at the very bottom ($Mo=1.37\times10^{-13}$), then water ($Mo=2.54\times10^{-11}$ at $20^\circ\,$C), then silicone oils T0 ($Mo=1.8\times10^{-10}$), T2 ($Mo=1.6\times10^{-8}$) and T5 ($Mo=6.2\times10^{-7}$) from bottom to top (see e.g. \cite{Zenit2008} for the characteristics of these oils). 
}
  \vspace{-86mm}\hspace{-48mm}$(a)$\hspace{65mm}$(b)$\\

  \vspace{38mm}\hspace{-117mm}
  $(c)$
   \vspace{37mm}
\label{f3.1.1}
\end{figure}


Figure~\ref{f3.1.1} summarizes the various interaction scenarios observed in the parameter range covered by the present investigation, $30<Ga\leq90$, $Bo\leq1.0$. The results from ZNM at $Ga=30$ are also reported in figure~\ref{f3.1.1}$(a)$ to serve as a reference. At this moderate $Ga$, 
the Asymmetric Side Escape (ASE) scenario is observed up to $Bo\approx0.45$, beyond which the two bubbles collide and coalesce. At $Ga = 40$ and $50$, the ASE scenario is still present, but only for Bond numbers less than a critical value $Bo=Bo_{o} \approx 0.2$. Beyond this threshold, once the trailing bubble (hereinafter abbreviated as TB) has drifted out of the wake of the leading bubble (abbreviated as LB), both bubbles exhibit oscillatory paths. For Bond numbers just beyond $Bo_{o}$ ($Bo=0.3-0.4$ in figure~\ref{f3.1.1}$(a)$), a first oscillatory regime is observed, with the two bubbles zigzagging virtually in the same plane and independently from each other (left panel in figure ~\ref{f3.1.1}$(c)$). We call it the Coplanar Independent Zigzagging regime (hereinafter abbreviated as CIZ). At $Ga=40$ this regime subsists until a second critical Bond number $Bo=Bo_{c}\approx0.5$ beyond which the bubbles collide. Following the remark in \S\,\ref{sec3}, we stopped the corresponding simulations when the two bubbles were very close to each other and did not attempt to capture the next steps of their dynamics. Increasing the Galilei number to $Ga=50$, two new regimes  are identified in between the CIZ and collision regimes. First, for $Bo=0.4$, the two bubbles still rise independently but their paths stand in two distinct preferential planes, with only tiny excursions of each bubble out of each of them. For this reason, we qualify the corresponding regime as Non-coplanar Independent Zigzagging (hereinafter abbreviated as NIZ). Clearly the CIZ-NIZ transition corresponds to a loss of the planar symmetry. 
Then, increasing the Bond number to $0.5$ reveals another type of evolution, which we identify as the Interacting Flattened Spiralling (IFS) regime. Here again, the two bubbles do not rise within the same plane after some time but, as will be made clear later, the motion of the TB continues to be deeply influenced by that of the LB during a significant part of the rise, if not throughout it. The two paths look like very flattened spirals, with the midplane of the TB path slowly rotating in such a way that in several cases the two midplanes are almost orthogonal at the end of the simulation (right panel in figure ~\ref{f3.1.1}$(c)$). Last, the two bubbles collide beyond $Bo_{c} \approx0.7$.  For higher Galilei numbers ($Ga\geq70$), the succession of the NIZ, IFS and collision regimes is observed beyond $Bo=0.1$. In contrast, no CIZ regime is encountered for smaller Bond numbers. In that range, instead of performing large-amplitude planar zigzags after the TB has moved out from the LB wake, the two bubbles rise independently from each other with only small-amplitude erratic horizontal excursions. This defines the Small-amplitude Independent Erratic (SIE) regime.

To properly interpret figure~\ref{f3.1.1}, it is important to keep in mind what the evolution of an isolated bubble with the same parameters would be. The critical curve corresponding to the onset of path instability for an isolated bubble as determined by CL16 is reported in figure~\ref{f3.1.1}$(a)$ (red dash-dotted line). This curve is found to coincide with the transition from the ASE scenario to the CIZ and SIE regimes up to $Ga=70$ (the curve is uncertain for $Bo\lesssim0.05$ and $Ga\gtrsim70$, as only Bond numbers of $\mathcal{O}(0.1)$ or larger were considered in that $Ga$-range in CL16). This is a clear indication that the CIZ and SIE regimes (and \textit{a fortiori} the NIZ and IFS regimes), correspond to conditions under which the rectilinear path of each bubble taken alone would be unstable. When the two bubbles are released in line and the Bond number is such that $Bo<Bo_c(Ga)$, the lateral drift of the TB triggers the path instability, giving rise to one of the above five scenarios. The central issue to be examined below is then to determine if and up to which point the subsequent evolution of the two bubbles is similar to that they would follow if rising alone, or if they rather continue to interact and the observed non-straight paths are still influenced by this interaction. 

\subsection{CIZ regime}\label{sec3.2}
\begin{figure}
\vspace{6mm}
  \centerline{\includegraphics[width=0.97\textwidth]{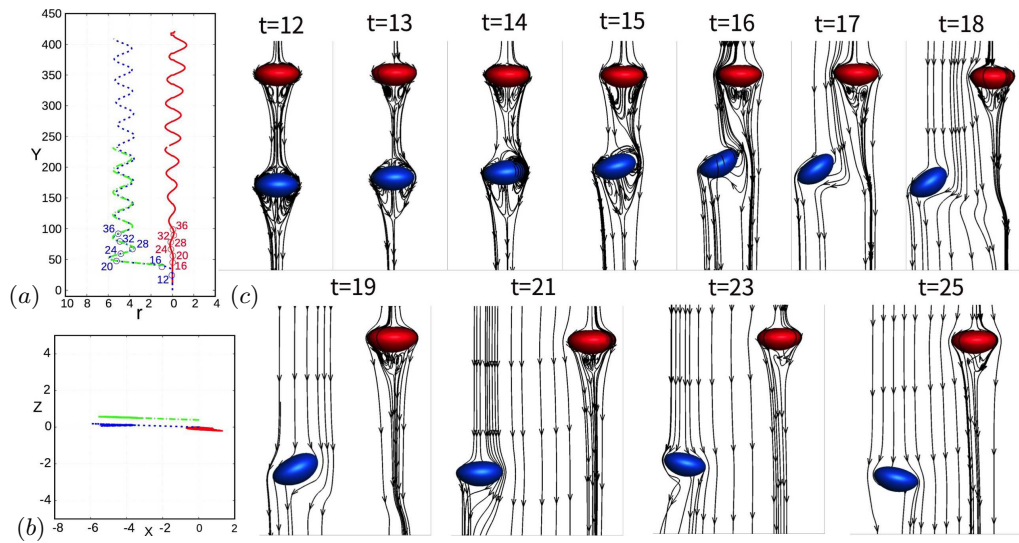}}
  \caption{Paths of a bubble pair with $(Ga,Bo) = (50,0.3$) following a CIZ evolution. $(a)$ Side view, with $r$ the transverse distance to the initial path and numbers indicating the  time instant at the corresponding position; $(b)$ bottom view; $(c)$ snapshots of the streamlines past the tandem in the reference frame of the LB. In $(a)$ and $(b)$, the red solid and blue dashed lines refer to the LB and TB, respectively, while the green dash-dotted line corresponds to the path of the corresponding isolated bubble.}
  \vspace{-57mm}\hspace{-1mm}$(a)$\hspace{25mm}$(c)$\\
  \hspace{-5mm}\\
  \vspace{20mm}\\
  $(b)$
   \vspace{19mm}
\label{f3.2.1}
\end{figure}

Although this regime is only encountered within a narrow range of parameters, it is the one with the simplest characteristics. This is why we discuss it first, based on the parameter set $(Ga, Bo) = (50, 0.3)$. An isolated bubble with these parameters follows a planar zigzagging path and its terminal Reynolds number and aspect ratio are $Re\approx125$ and $\chi \approx 2.1$, respectively (CL16). The evolution of the corresponding paths and streamlines for the bubble pair are shown in figure~\ref{f3.2.1}. Not surprisingly, the front view of the paths (panel $(a)$) reveals that the initial in-line configuration breaks down in the very asymmetric manner typical of the ASE scenario ($12\lesssim t\lesssim20$). Then the TB immediately starts to follow a large-amplitude planar zigzagging path. In contrast, the LB slowly transitions to a similar path and reaches a nearly-saturated stage only near the end of the computation. In this `final' stage, the two oscillatory paths have almost identical frequencies and amplitudes, and these quantities are similar to those found with an isolated bubble (green line). These features suggest that the two bubbles rise independently of each other. 
The bottom view of the paths (panel $(b)$) shows that, up to tiny deviations, the entire sequence takes place within a single vertical plane. This plane is that defined by the initial ASE interaction, i.e. the vertical path of the LB and the inclined path of the TB during its lateral escape at early times ($t\lesssim16$). This lateral drift provides a finite-amplitude disturbance that triggers the path instability of both bubbles. However the corresponding asymmetry is much weaker in the neighbouring of the LB than around the TB, which results in a much slower development of the oscillatory motion of the former. This difference in the magnitude of the flow asymmetry about the two bubbles may be appreciated by examining the distortion of the streamlines and the inclination of the bubble minor axis at $t\approx18-19$ (panel $(c)$). Remarkably, under present conditions, the path instability mechanism conserves the planar symmetry dictated by this early evolution of the system. As we shall see later, this is an exception rather than a general rule.\\
\begin{figure}
\vspace{6mm}
  \centerline{\includegraphics[width=0.98\textwidth]{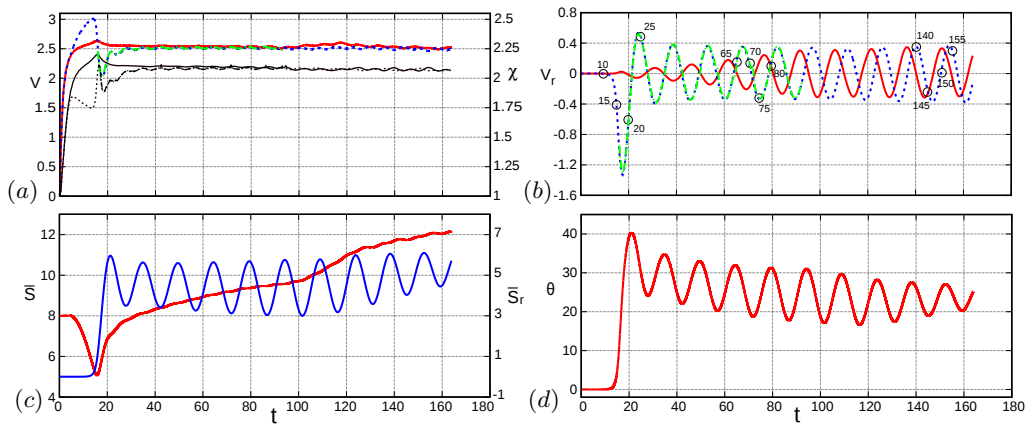}}
  \caption{Evolution of several characteristics of a bubble pair with $(Ga,Bo) = (50,0.3$). $(a)$ Rising speed (colored lines, left axis) and aspect ratio (black lines, right axis) of the LB (solid lines) and TB (dashed lines); $(b)$ same for the horizontal velocity component (numbers indicate the corresponding time instant); $(c)$ vertical (red line, left axis) and transverse (blue line, right axis) separations; $(d)$ inclination of the line of centres with respect to the vertical. In $(a)-(b)$ the color code is similar to that in figure \ref{f3.2.1}.}
   \vspace{-56mm}\hspace{-1mm}$(a)$\hspace{64mm}$(b)$\\
  \hspace{-5mm}\\
  \vspace{17mm}\\
  $(c)$\hspace{64mm}$(d)$\\
   \vspace{20mm}
\label{f3.2.2}
\end{figure}
\indent The evolution of some characteristics of the motion is displayed in figure~\ref{f3.2.2}.
The early evolution of the rising speed (panel $(a)$, left axis), is similar to that observed in the ASE scenario detailed in ZNM. That is, once the TB enters the wake of the LB, it is strongly accelerated by the corresponding `sheltering' effect, while the rising speed of the LB only experiences a slight increase resulting from the reduction of velocity gradients in its wake caused by the flow at the front of the TB ($4 \lesssim t \lesssim 13$). As a result, the vertical separation decreases sharply (panel $(c)$) and the aspect ratio of the TB (panel $(a)$, right axis) drops dramatically because of the suction induced by the low pressure at the back of the LB. 
Then, for the reasons discussed in ZNM, the in-line arrangement becomes unstable and the TB starts to drift laterally, as the records of the transverse separation $\overline{S}_r$ (panel $(c)$) and transverse velocity (panel $(b)$) indicate. Owing to this lateral drift, part of the potential energy of the TB is used to `feed' its horizontal motion, making its rising speed, $V_{TB}$, drop dramatically and become even lower than that of the LB, $V_{LB}$, which in turn results in a re-increase of the vertical separation beyond $t\approx15$. \\
\indent At $t \approx 19$, the TB has left the wake of the LB and starts zigzagging in a vertical plane. The first period of this oscillating motion is still influenced by the large-amplitude disturbance provided by the initial lateral drift. However this influence quickly fades away and the lateral excursions reach a saturated periodic state at $t\approx35$. Beyond this point, the  TB path is characterized by a rising speed $V_{TB}\approx2.5$, a dimensionless frequency $\overline{f}\approx0.073$ and a crest-to-crest amplitude $\overline{a}\approx1.75$ (figure \ref{f3.2.1}$(a)$), with maximum horizontal velocities (figure \ref{f3.2.2}$(b)$) of approximately $0.15V_{TB}$. These characteristics superimpose onto those of the corresponding isolated bubble, confirming that the LB no longer influences the TB. The oscillating component of the LB motion follows a strikingly different evolution. Indeed, it develops much more gradually and figure \ref{f3.2.2}$(b)$ shows that the oscillations of its lateral velocity, $V_{r_{LB}}$, still have a slightly smaller amplitude than those of $V_{r_{TB}}$ at $t\approx160$. Because of this small difference, a slightly larger fraction of the potential energy of the LB is converted into the kinetic energy associated with its rise, making $V_{LB}$ be still slightly larger than $V_{TB}$, as figure \ref{f3.2.2}$(a)$ confirms.  This is the origin of the gradual increase of the vertical separation still present at the end of the sequence in figure \ref{f3.2.2}$(c)$. The final geometry of the tandem may be anticipated from the late stages displayed in figures \ref{f3.2.2}$(c)-(d)$: the average transverse separation will be close to $5.5$, a distance at which the interaction between the two bubbles is extremely weak  at the current Reynolds number according to the predictions of \cite{hallez2011interaction} for spherical bubbles, and the average inclination of the line of centres will be approximately $24^\circ$. Due to the slight increase in the frequency of the LB path oscillations as they grow, the phase shift between the two paths has decreased continuously during the transient (figure \ref{f3.2.2}$(b)$). However, there is \textit{a priori} no reason for this phase shift to vanish eventually, given the quasi-independence of the two paths. Consequently, both $\overline{S}_r$ and $\theta$ will continue to oscillate slightly about their mean value. \\
\begin{figure}
\vspace{5mm}
  \centerline{\includegraphics[width=0.98\textwidth]{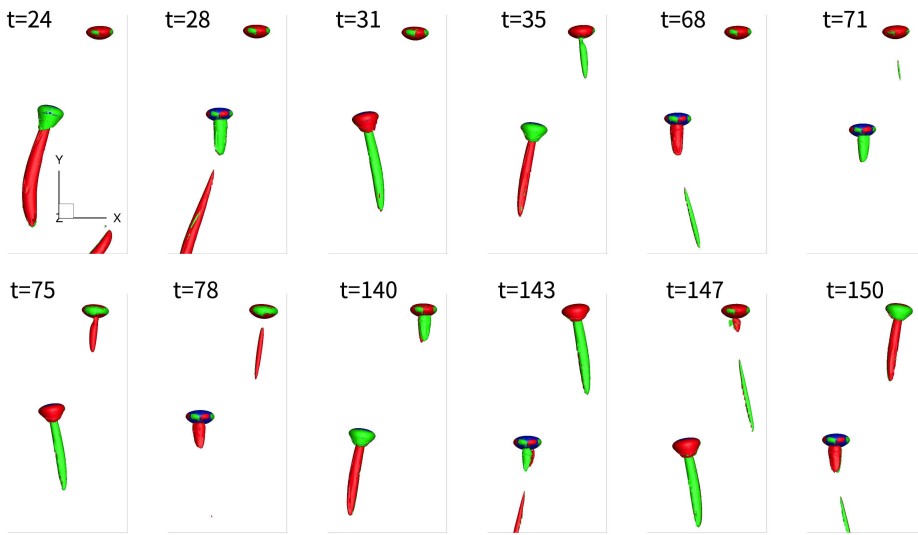}}
  \caption{Successive snapshots of the iso-surface $\omega_y = -0.5$ (green) and $\omega_y = 0.5$ (red) during the rise of a bubble pair with $(Ga,Bo) = (50,0.3$). }
\label{f3.2.3}
\end{figure}\indent 
Another perspective into the dynamics of the tandem is provided by the evolution of the streamwise vorticity distribution in the wake of the two bubbles. Indeed, a spheroidal bubble rising in a straight vertical line having an axisymmetric wake, vorticity is purely azimuthal in this case and has no streamwise, i.e. vertical, component. 
Successive snapshots of a selected iso-contour of the vertical vorticity, $\omega_y=\pm0.5$, are plotted in figure~\ref{f3.2.3}. Two vortex threads in which $\omega_y$ periodically changes sign emerge in the TB wake right after its initial lateral drift. In contrast, no structure corresponding to the selected $|\omega_y|$-level is found in the LB wake until $t\approx60$. This finding confirms that only the TB performs large-amplitude lateral oscillations at earlier times, since such oscillations directly result from the lift force originating in the double-threaded streamwise vortices. Then the size of the vortex threads past the LB increases continuously until $t\approx140$, beyond which the two pairs of threads reach a similar size, as could be anticipated from figure \ref{f3.2.2}$(b)$.
 It may be noted that, throughout their nearly-parallel rise, the separation between the two bubbles is large enough to avoid each of them to hit the pair of threads emitted by the other. This confirms that in the present regime only minimal interactions subsist between the two bubbles after the initial ASE stage. 

\subsection{NIZ regime}\label{sec3.3}

\begin{figure}
\vspace{6mm}
  \centerline{\includegraphics[width=0.98\textwidth]{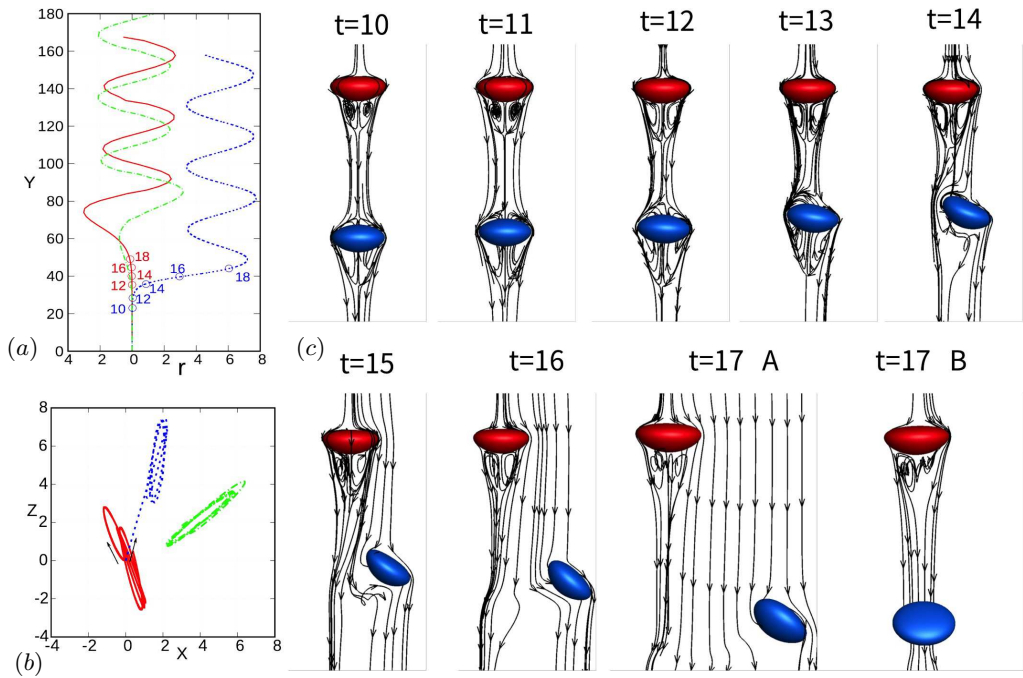}}
  \caption{Paths of a bubble pair with $(Ga,Bo) = (70,0.3)$ following a NIZ evolution. $(a)$ Side view, with $r$ the transverse distance to the initial path and numbers indicating the  time instant at the corresponding position; $(b)$ top view; $(c)$ snapshots of the three-dimensional streamlines past the two bubbles in the reference frame of the LB. At $t = 17$, streamlines are shown in two perpendicular vertical planes, plane A containing the initial vertical path and that followed by the TB during its first lateral drift. In $(a)-(b)$, the color code is similar to that in figure \ref{f3.2.1}; the side view in $(a)$ was obtained by considering the path of each bubble in its own plane and juxtaposing the three paths in the same plane.} 
   \vspace{-75mm}\hspace{-1mm}$(a)$\hspace{34mm}$(c)$\\
  \hspace{-5mm}\\
  \vspace{31mm}\\
  $(b)$
   \vspace{28mm}
\label{f3.3.1}
\end{figure}

\begin{figure}
 \vspace{6mm}
\centerline{\includegraphics[width=0.98\textwidth]{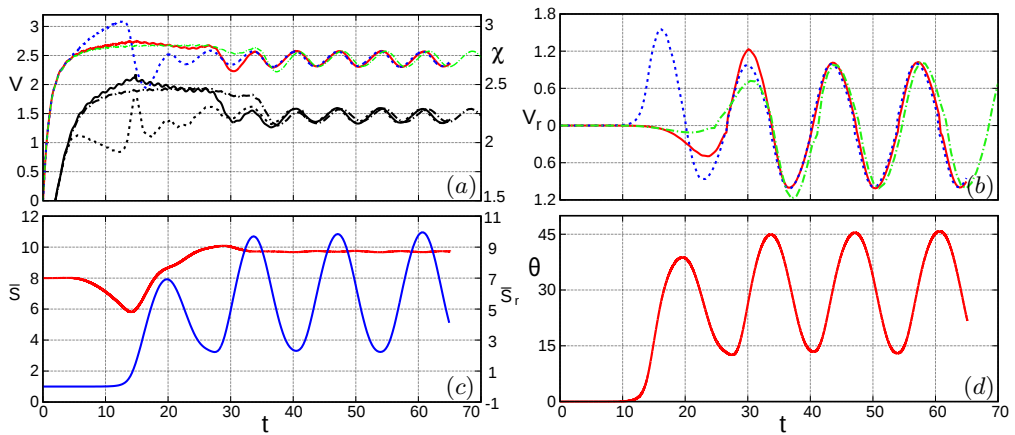}}
  \caption{Evolution of several characteristics of a bubble pair with $(Ga,Bo) = (70,0.3$). $(a)$ Rising speed (colored lines, left axis) and aspect ratio (black lines, right axis) of the LB (solid lines), TB (dashed lines) and isolated bubble (dash-dotted lines); $(b)$ same for the horizontal velocity component; $(c)$ vertical (red line, left axis) and transverse (blue line, right axis) separations; $(d)$ inclination of the line of centres with respect to the vertical. 
  }
   \vspace{-54.5mm}\hspace{59mm}$(a)$\hspace{64.5mm}$(b)$\\
   
  \vspace{19.5mm}\hspace{59mm}$(c)$\hspace{64.5mm}$(d)$\\
   \vspace{21mm}
\label{f3.3.3}
\end{figure}

We now increase to ratio of inertial to viscous effects by examining the results obtained with $(Ga, Bo) = (70, 0.3)$ which, according to the phase diagram of figure \ref{f3.1.1}, belongs to the NIZ regime. Under such conditions, an isolated bubble follows a flattened spiralling motion, and its terminal rising speed and aspect ratios are $Re\approx168$ and $\chi \approx2.2$, respectively (CL16). 
The major two differences in the evolution of this bubble pair compared to the previous one may be appreciated from figures \ref{f3.3.1}$(a)-(b)$. First, both bubbles now follow large-amplitude oscillatory paths soon after the initial ASE interaction is completed. Second, these oscillatory motions take place in two distinct vertical planes.  These two figures also show 
that both paths have frequencies and amplitudes very similar to those of the corresponding isolated bubble, which again indicates that the two bubbles rise almost independently beyond the initial ASE stage.\\
 \indent Figure~\ref{f3.3.1}$(c)$ displays the evolution of the bubbles shape and orientation, together with that of the streamlines pattern. The comparison with figure~\ref{f3.2.1}$(c)$ in the early stage, for instance at $t=12$, is enlightening. Clearly, the front part of the LB is significantly flatter in figure~\ref{f3.3.1}$(c)$, and the rear part is slightly more rounded. Also, the standing eddy at the back of the same bubble is much larger. That an increase in $Ga$ (or $Re$) increases the fore-aft asymmetry of an isolated rising bubble in the above way is a well-documented effect \citep{Ryskin1984,Zenit2008}. It is directly related to the influence of the shear-free condition on the pressure distribution in the liquid at the interface, i.e. to the presence of a boundary layer around the bubble. Since the flattening of the front is stronger than the rounding of the rear, increasing $Ga$ makes the bubble aspect ratio increase. This in turn has a direct influence on the magnitude of the azimuthal vorticity generated at the gas-liquid interface, since the maximum of this surface vorticity (reached near the buble equator) is known to vary as $\chi^{8/3}$ for large $\chi$ \citep{magnaudet2007wake}. This increase in the amount of vorticity produced at the surface of the LB is the direct cause of the larger size of the standing eddy noticed above. It is also the cause of the vigorous path instability revealed by the evolution of the LB path. Indeed, it has been established that wake instability past a perfectly spheroidal bubble sets in when the aspect ratio exceeds a threshold $\chi_c\approx2.2$ \citep{magnaudet2007wake}. 
 As the record in figure \ref{f3.3.3}$(a)$ indicates, with $(Ga, Bo) = (70, 0.3)$ the aspect ratio of the LB stays beyond $2.25$ from $t\approx5$ to $t\approx30$. Therefore, the conditions required for the wake (hence the path) of this bubble to become unstable are fulfilled, which explains why it has already performed the first half of a large-amplitude zigzag at $t=30$. In comparison, with $(Ga, Bo) = (50, 0.3)$, figure \ref{f3.2.2}$(a)$ indicates that the aspect ratio of the LB exceeds $2.2$ only during a short transient near $t=15$. This is why the lateral excursions of this bubble grow much more slowly.\\
 \indent Panels $(b)-(d)$ in figure~\ref{f3.3.3} confirm that the dynamics of the tandem have reached a fully-developed state soon after $t=30$. Indeed, the vertical separation stabilizes at $\overline{S}\approx9.7$ and no longer varies. Owing to the phase shift between the two paths revealed by figure \ref{f3.3.1}$(a)$, the horizontal separation $\overline{S}_r$ experiences periodic large-amplitude oscillations with a frequency $\overline{f}\approx0.074$. These oscillations make $\overline{S}_r$ vary from $2.25$ to $10$, which in turn results in an inclination of the tandem varying from $13^\circ$ to $45^\circ$.  As figure~\ref{f3.3.3}$(a)$ shows, the rising speed and aspect ratio of each bubble experience significant oscillations during this fully-developed stage. The frequency of these oscillations is twice that of the other quantities, the rising speed reaching its maximum twice during a complete zigzag. The oscillations of the aspect ratio are enslaved to those of the rising speed, being driven by the instantaneous value of the Weber number. A noticeable feature of the transition from the initial transient to the fully-developed state is the significant decrease in the rising speed of the LB (similar to that of the isolated bubble), from $V_{LB}\approx2.65$ for $t\lesssim28$ to an average value close to $2.45$ at later times. As discussed in CL16, this reduction results on the one hand from the increased dissipation in the wake associated with the double-threaded trailing vortices, and on the other hand from the part of the bubble potential energy spent to `feed' its lateral excursions rather than its rise.\\
 \indent 
As pointed out before, the top view of the path (figure \ref{f3.3.1}$(b)$) reveals that the LB oscillates in a plane (say plane B) which does not coincide with that selected by the TB during its initial lateral drift (say plane A). An early indication of this angular splitting is provided by the two perpendicular views at $t=17$ in figure \ref{f3.3.1}$(c)$. Indeed, the flow at the back of the LB is not symmetric in plane B, nor is the projected bubble shape. Consequently, the plane within which the flow past the LB is symmetric (if any) is neither A nor B, and this bubble is just starting to oscillate in a vertical plane with an intermediate orientation. Given the above discussion, the reason for this behaviour is obvious: the oblateness of the LB being beyond the wake instability threshold, non-axisymmetric disturbances round this bubble develop vigorously and are almost uncorrelated with those generated by the TB drift. Because of this, the system does not preserve any planar symmetry.
\subsection{SIE regime}\label{sec3.4.0}
\begin{figure}
\vspace{6mm}
  \centerline{\includegraphics[width=0.98\textwidth]{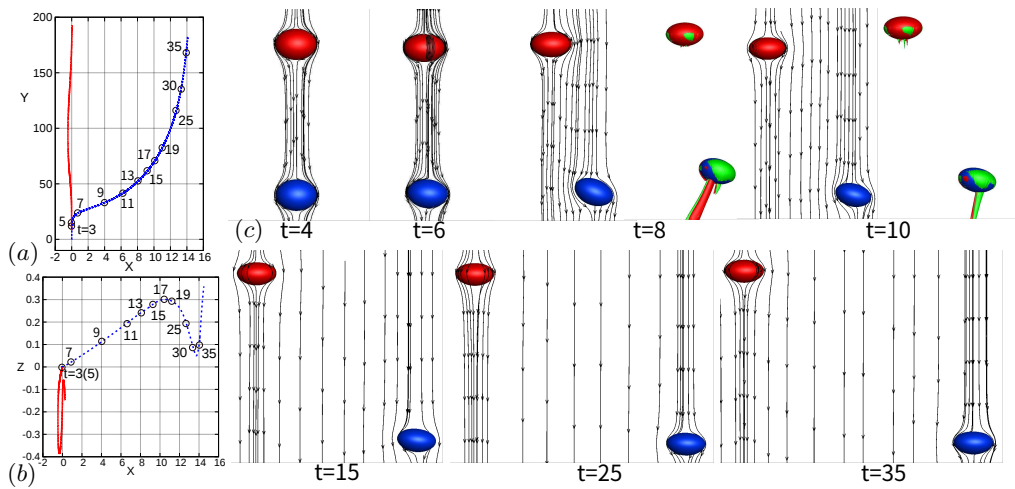}}
  \caption{Paths of a bubble pair in the SIE regime ($Ga=90,Bo=0.05$). $(a)$ Side view, with numbers indicating the time instant at the corresponding position; $(b)$ top view; $(c)$ snapshots of the three-dimensional streamlines past the two bubbles in the reference frame of the LB. The wake structure at $t=8$ and $10$ is also shown using the iso-surfaces $\omega_y=\pm3$. In $(a)-(b)$, the color and line codes are similar to those of figure \ref{f3.2.1}.} 
   \vspace{-55mm}\hspace{30mm}$(c)$\\
   
   \vspace{-4mm}
   \hspace{0mm}$(a)$\\
   
  \hspace{-18mm}\\
  \vspace{15mm}\\
  $(b)$
   \vspace{17mm}
\label{SIE1}
\end{figure}
For $Ga\geq70$ and $Bo\leq0.1$, a specific regime takes place. Here, once the TB has drifted out of the LB wake, the two bubbles do not follow large-amplitude zigzagging or spiralling paths, nor do they rise strictly in a straight line. Figure \ref{SIE1} shows a typical example of this regime, corresponding to $Ga=90$ and $Bo=0.05$. As panel $(a)$ reveals, the LB almost rises vertically throughout its ascent, with however some tiny meandering. On the other hand, after having started its lateral drift very early through the usual ASE mechanism, the TB continues to drift gradually well after it has moved out of the LB wake. In figure \ref{SIE1}$(b)$ it is noticed that the TB keeps on rising within a vertical plane during some time ($t\lesssim17$), before its path becomes three-dimensional. Then the horizontal excursions of this bubble exhibit erratic orientations, as do those of the LB from the very beginning of its ascent. These horizontal motions remain of small amplitude, typically some tenths of the bubble radius, to be compared to $2-4$ radii in figures \ref{f3.2.1} and \ref{f3.3.1}. Clearly, the two bubbles no longer interact beyond the initial ASE stage. The above characteristics are shared by all bubble pairs standing in the SIE regime.\\  \begin{figure}
 \vspace{6mm}
\centerline{\includegraphics[width=0.98\textwidth]{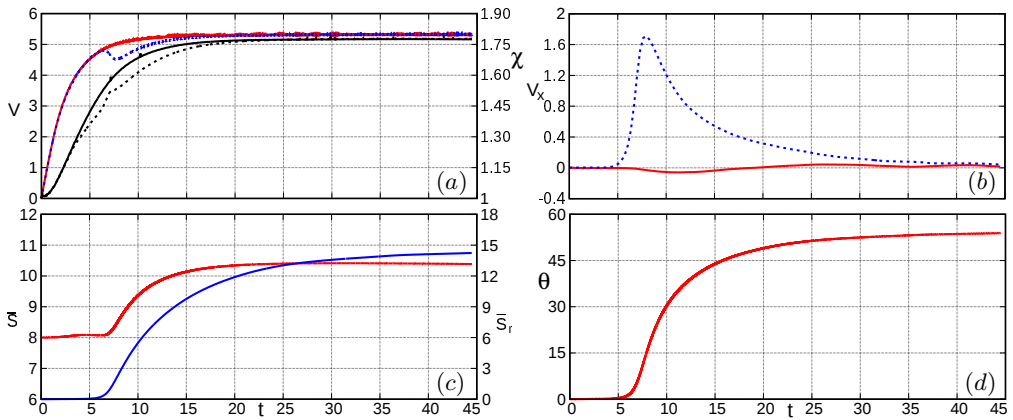}}
  \caption{Evolution of several characteristics of a bubble pair with $(Ga,Bo) = (90,0.05$). $(a)$ Rising speed (colored lines, left axis) and aspect ratio (black lines, right axis) of the LB (solid lines) and TB (dashed lines); $(b)$ same for the horizontal velocity component; $(c)$ vertical (red line, left axis) and transverse (blue line, right axis) separations; $(d)$ inclination of the line of centres with respect to the vertical.}
   \vspace{-53mm}\hspace{58mm}$(a)$\hspace{66mm}$(b)$\\
   
  \vspace{19.5mm}\hspace{58mm}$(c)$\hspace{66mm}$(d)$\\
   \vspace{17mm}
\label{SIE2}
\end{figure}\indent 
A remarkable feature in the streamline patterns displayed in figure \ref{SIE1}$(c)$ is that no standing eddy exists past the two bubbles. Their final aspect ratio is close to $1.75$ (figures \ref{f3.1.1}$(b)$ and \ref{SIE2}$(a)$), an oblateness for which the numerical results of \cite{blanco1995structure} for an isolated bubble indicate that a standing eddy exists only if the Reynolds number is less than $160$. Here the final Reynolds number of the two bubbles is almost three times larger ($Re\approx470$, see figure \ref{f3.1.1}$(b)$), implying a significantly lower accumulation of vorticity at the back of the bubble. Consequently, the absence of standing eddies in figure \ref{SIE1}$(c)$ is no surprise. What may sound more surprising is that the system exhibits path instability (albeit with weak manifestations) while the wake instability past a fixed spheroidal bubble only takes place if the aspect ratio is beyond $\chi_c\approx2.2$ \citep{magnaudet2007wake}. Similar small-amplitude chaotic paths were identified in CL16 in the same $(Ga,Bo)$ range as the present SIE regime. As discussed in that reference, the reason why the path of a bubble may be unstable while its wake is still intrinsically stable is that freely-moving bubbles must satisfy overall constraints related to the zero-torque and constant-force conditions. These constraints are responsible for the existence of specific instability modes, which may under certain circumstances become unstable before (i.e. at lower $\chi$) than those associated with the wake instability. Possible shape oscillations play no role in this phenomenon, as it also exists (and is better documented) for rigid bodies. For instance, numerical simulation \citep{Auguste2013} and linear stability analysis \citep{Tchoufag2014b} reveal that, for certain solid-to-fluid density ratios, path instability of freely-falling disks occurs at a critical Reynolds number more than three times smaller than the threshold of wake instability past a disk held fixed at normal incidence in a uniform stream.\\
\indent Figure \ref{SIE2} helps understand the reason and consequences of the unexpectedly long drift of the TB. In particular, panel $(a)$ reveals that, while the rising speed of the LB has reached its final value at $t\approx13$, its aspect ratio still increases significantly until $t\approx20$. This is an indication that the bubbles have not yet reached their final shape during the stage when they interact. The reason for this delay is that the rate at which the bubble shape adapts to the dynamical conditions is limited by viscous diffusion. As the corresponding characteristic time is very long under such highly inertial conditions (the viscous time scale being proportional to $Re$), the aspect ratio $\chi(t)$ at a given time is lower than what the quasi-static approximation of \cite{Moore1965} based on the Weber number $We(t)$ built with the rising speed $V(t)$  would predict. A similar argument applies to the evolution of the TB wake. During the time the TB drifts across the flow region disturbed by the presence of the LB (say $t\lesssim10$ according to figure \ref{SIE1}$(c)$), its wake exhibits the classical double-threaded structure resulting from the ambient shear created by this disturbance (see the snapshots at $t=8$ and $10$). Once the TB has reached the flow region left undisturbed by the LB, its wake tends to recover its axial symmetry since it is intrinsically stable. However the return to this symmetry is governed by the disappearance of the trailing vortices and the reorientation of the bubble whose minor axis must realign with the vertical. Viscous mechanisms playing a central role in both processes, the return to an axisymmetric flow structure past the TB is far from instantaneous, as the slow decay of the transverse velocity in figure \ref{SIE2}$(b)$ and the nonzero inclination of the bubble equator at $t=15$ in figure \ref{SIE1}$(c)$ confirm. This is why the TB continues to drift laterally until the uppermost position reached in the computation. A direct consequence of this sustained drift is that the long-term lateral separation reaches much larger values than in the previous regimes ($\overline{S}_r\approx15$ according to figure \ref{SIE2}$(c)$, to be compared with average values in the range $5-6$ in the CIZ and NIZ regimes). This is turn translates into a significantly larger final inclination of the bubble pair, close to $55^\circ$ (figure \ref{SIE2}$(d)$). In anticipation to the regime discussed below with the same $Ga$ but a ten times larger $Bo$ (see figure \ref{f3.4.11}$(c)$), one may also notice in figure \ref{SIE2}$(c)$ that the vertical separation does not drop during the initial period when the TB stands in the LB wake. Indeed, such low-$Bo$ bubbles being only moderately deformed and their rise Reynolds number being very large even at this early stage, their wakes remain thin, making the attractive influence of the LB too weak to reduce significantly $\overline{S}$ below its initial value.

\subsection{IFS regime}\label{sec3.4}
\begin{figure}
\vspace{6mm}
  \centerline{\includegraphics[width=0.98\textwidth]{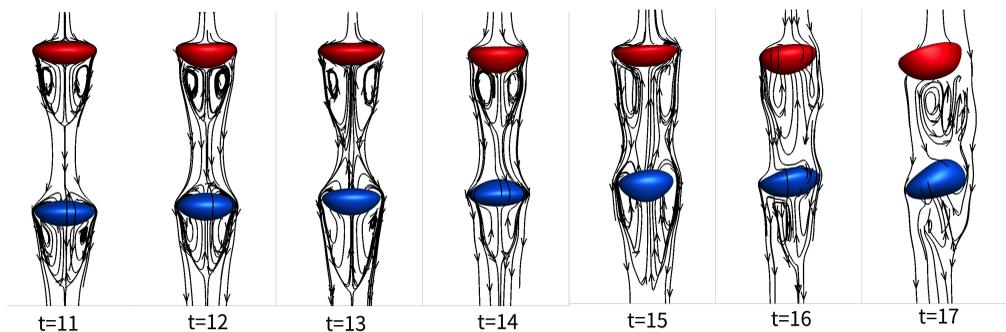}}
  \caption{Three-dimensional streamlines past a bubble pair with $(Ga, Bo) = (90, 0.5)$ during the early stage of their interaction (in the LB reference frame).}
\label{f3.4.2}
\end{figure}
\begin{figure}
\vspace{6mm}
  \centerline{\includegraphics[width=0.4\textwidth]{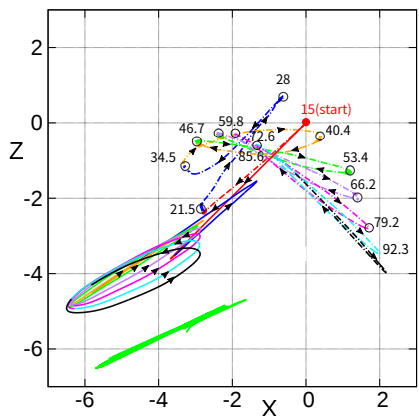}}
  \caption{Evolution of the horizontal trace of the paths for a pair of bubbles with $(Ga,Bo)=(90,0.5)$; the solid and dash-dotted lines refer to the LB and TB, respectively. Successive periods of the path are coloured in red/blue/orange/green/purple/pink/cyan/black; numbers along the TB path indicate the corresponding time instant. The green trace in the bottom part is that of the corresponding isolated bubble, the initial ($X,Z)$ position of which has been shifted for readability.}
  \vspace{-3mm}
\label{f3.4.1}
\end{figure}

We now focus on the set of parameters $(Ga, Bo) = (90, 0.5)$. With these parameters, an isolated bubble follows a flattened spiralling path (CL16), with final characteristics $Re \approx 175$ and $\chi \approx 2.6$. Figure \ref{f3.4.2} shows some streamlines past the two bubbles during their initial in-line rise. Owing to the large curvature of the LB in the equatorial region, a large standing eddy develops behind it. Asymmetries in the streamlines pattern become visible at $t\approx13$. At this moment, the standing eddy at the back of the LB and the disturbance flow past the TB are directly connected, since the closure zone of the former almost coincides with the upper end of the closed streamlines ahead of the TB. Therefore, no intermediate region with a nearly-parallel flow subsists in between the two bubbles, unlike the situation prevailing at earlier times. For this reason, the two bubbles are tightly coupled and any tilting of one of them tilts the other in the same direction, as the last two snapshots in the series confirm. This is why they both start to drift in the same direction. Moreover, the flow asymmetry associated with this lateral motion makes both of them take a pronounced egg-like shape, with the pointed end directed toward the direction of the drift. That the two bubbles initially drift in the same direction is strikingly different from what was observed in the previous three regimes, which all involve an initial ASE-type stage.\\
\indent The bottom view of the paths at larger times is displayed in figure~\ref{f3.4.1}, the initial (red) stage corresponding to the time interval during which the two bubbles drift in the same direction. The path of the corresponding isolated bubble is also shown to serve as a reference (green trace in the bottom part of the figure). It reveals that the bubble follows an almost perfect planar zigzagging motion. 
In the case of the bubble pair, after a transitional (blue) stage, say beyond $t\gtrsim35$, the LB first follows a nearly planar zigzagging path (orange stage), while the TB starts to describe a flattening spiral with a $5:1$ aspect ratio. Beyond this first complete oscillation, the two bubbles follow significantly different evolutions. The initial planar zigzagging path of the LB turns gradually into a flattened spiralling motion whose aspect ratio decreases over time and is close to $4.5:1$ at the end of the simulation. Meanwhile, the corresponding path experiences a weak clockwise precession, the mid-plane of the oscillations having rotated by approximately $10^\circ$ over the five complete oscillations covered by the simulation. Conversely, the path oscillations of the TB become more and more two-dimensional: while the first (orange) flattened spiral beyond $t=35$ has a $6.5:1$ aspect ratio, the last (black) oscillation is virtually a planar zigzag. In the meantime, the mid-plane of the oscillations has rotated clockwise by approximately $60^\circ$, revealing a precession six times faster than that of the LB. As a result, the two horizontal traces of the paths which were initially nearly aligned make and angle close to $80^\circ$ at the end of the simulation. 
\\
\begin{figure}
\vspace{3mm}
  \centerline{\includegraphics[width=0.98\textwidth]{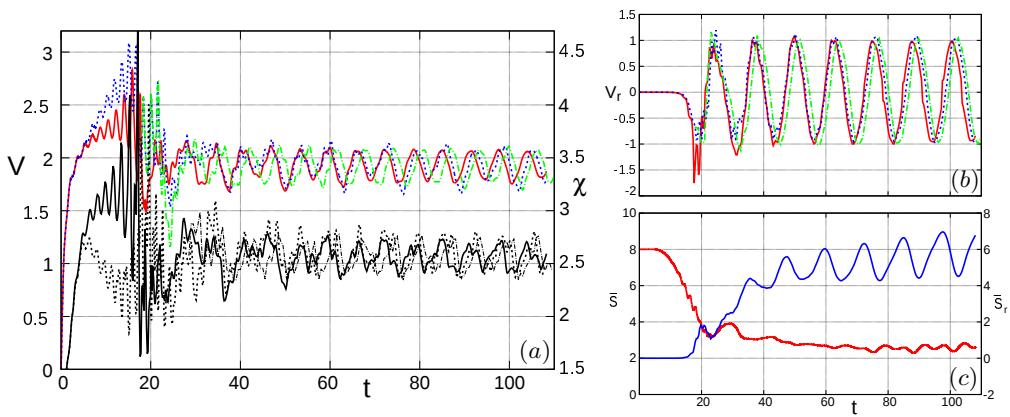}}
  \caption{Evolution of several characteristics of a bubble pair with $(Ga,Bo)=(90,0.5)$. $(a)$: rising speed (colored lines, left axis) and aspect ratio (black lines, right axis) of the LB (solid lines) and TB (dashed lines); $(b)$ same for the horizontal velocity component; $(c)$ vertical (red line, left axis) and transverse (blue line, right axis) separations. In $(a)-(b)$, the color code is similar to that in figure \ref{f3.2.1}.}
    \vspace{-52.8mm}\hspace{126mm}$(b)$\\
    
  \vspace{15.5mm}\hspace{69mm}$(a)$\\
  
   \vspace{-3.5mm}\hspace{125.8mm}$(c)$\\
   \vspace{19mm}

\label{f3.4.11}
\end{figure}
\indent Figure \ref{f3.4.11}$(a)$ shows how the rising speed and aspect ratio of the two bubbles vary along their rise. Both quantities experience strong and rapid variations during the stage $15\lesssim t\lesssim25$ which corresponds to the setting up of the zigzagging/spiralling motions. Beyond this point, they reach a `developed' state characterized by well-defined mean values on which quasi-periodic oscillations with a dominant frequency $\overline{f}\approx0.156$ superimpose. These oscillations are far from sinusoidal and a series of higher harmonics with a significant amplitude is present. This is especially visible in the evolution of the aspect ratio which exhibits numerous spikes, reflecting the fact that the strong and asymmetric deformation of both bubbles involves a combination of modes. It is worth noting that the average rising speed of each bubble is close to $1.9$, to be compared with values close to $2.5$ fin the CIZ and NIZ regimes (see figures \ref{f3.2.2} and \ref{f3.3.3}), or even $5.3$ the case of the SIE regime (figure \ref{SIE2}). This is a direct consequence of the larger bubble deformation which implies stronger velocity gradients in the near-bubble flow, from which a larger drag results. Quite remarkably, the transverse velocities seen in figure \ref{f3.4.11}$(b)$ exhibit very similar evolutions, both in amplitude and frequency, despite the significant differences noticed in the record of the lateral motions. Differences are however visible during short specific stages, most notably near the extrema. \\
\indent The records of the horizontal and vertical separations in figure \ref{f3.4.11}$(c)$ reveal interesting features. In particular, apart from a small bump during the stage $23\lesssim t\lesssim33$, the vertical separation continuously decreases over time, finally stabilizing around $\overline{S}\approx2.6$. As a result, beyond $t\approx70$, the inclination of the line of centres (not shown) stabilizes around $\theta\approx65^\circ$ with $\pm5^\circ$ oscillations. This small final vertical separation is to be compared with the final values $\overline{S}\approx10$ or even larger found in figures \ref{f3.2.2}, \ref{f3.3.3} and \ref{SIE2}. Most of the decrease of $\overline{S}$ takes place before the horizontal oscillations fully develop. The reason for this is easily identified from figures \ref{f3.4.2} and \ref{f3.4.1}. Since the two bubbles drift laterally in the same direction during a significant time, the `shielding' effect of the LB lasts for a much longer time than in the previous examples, allowing the TB to get much closer to the LB before the zigzagging/spiralling oscillations fully develop. This makes the vertical separation decrease continuously up to $t\approx23$, when it reaches a first minimum close to $3.2$. The evolution of the horizontal separation is qualitatively similar to those observed in the CIZ and NIZ regimes. In particular the average separation is again close to $6$ at the end of the sequence. In contrast, the amplitude of the $\overline{S}_r$-oscillations is nearly three times less than in figure \ref{f3.3.3}$(c)$, a result of the gradual reduction of the lateral excursions of the LB. Both the average value and the oscillations of $\overline{S}_r$ are still growing at the end of the sequence, confirming that the two bubbles keep on following different horizontal dynamics.  \\
\begin{figure}
\vspace{6mm}
  \centerline{\includegraphics[width=0.98\textwidth]{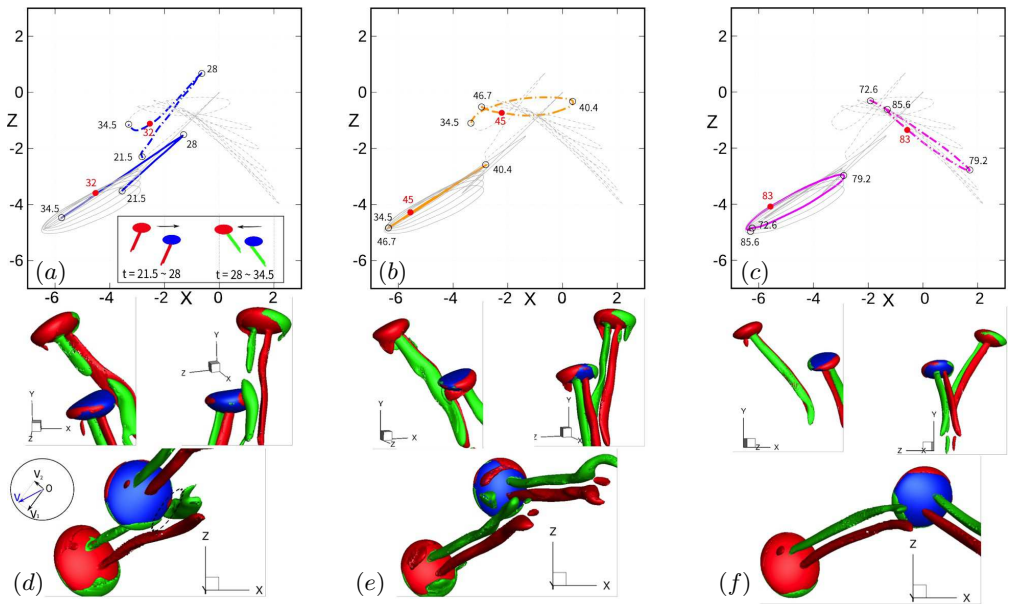}}
  \caption{Bubble-wake interactions during three successive stages of the spiralling/zigzagging motion of a bubble pair with $(Ga,Bo) =(90,0.5$). $(a)-(c)$: bottom view of the path for $21.5 < t < 34.5$, $34.5 < t < 46.7$ and $72.6 < t < 85.6$, respectively (the inset in $(a)$ shows the two successive arrangements of the bubble pair in the horizontal direction). $(d)-(f)$ iso-surfaces $\omega_y = +3$ (red) and $\omega_y =-3$ (green) of the vertical vorticity at $t = 32$, $45$ and $83$, respectively (the top row displays two orthogonal side views, while the bottom row shows the bottom view).}
  \vspace{-71mm}
   \hspace{5mm}$(a)$ \hspace{40mm}$(b)$\hspace{44mm}$(c)$\\
   
    \vspace{34mm} 
    \hspace{2mm}$(d)$ \hspace{40mm}$(e)$\hspace{44mm}$(f)$
     \vspace{27mm}
\label{f3.4.3}
\end{figure}
\indent Figure~\ref{f3.4.3} details the path and wake structure of the two bubbles during three selected stages of their rise. In each of them, the tandem first drifts toward the right with the TB ahead of the LB, as shown in the inset of panel $(a)$. Then the direction of the drift reverses and the LB takes the lead. 
In the first half of each stage, say $21.5 < t < 28$ in $(a)$, the LB wake hardly alters the motion of the TB which drifts ahead of it. In contrast, in the second half, say $28 < t < 34.5$ in $(a)$, the double-threaded vortex structure shed by the LB hits the TB, especially through the negative (green) thread, and thereby influences its path. In $(b)$ this interaction is responsible for the smaller radius of curvature of the TB path observed during the time interval $40.4 < t < 46.7$, a feature that initiates the clockwise precession characterizing the subsequent evolution of this path. 
This scenario repeats itself and the clockwise precession of the TB path increases every time the negative vortex thread hits the TB. 
However, as figure~\ref{f3.4.1} shows, this precession gradually makes the paths of the two bubble develop in orthogonal planes. This tends to keep the transverse distance between the two bubbles larger during their ascent, thus reducing the opportunities for the LB wake to directly influence the TB. 
\begin{figure}
  \centerline{\includegraphics[width=0.98\textwidth]{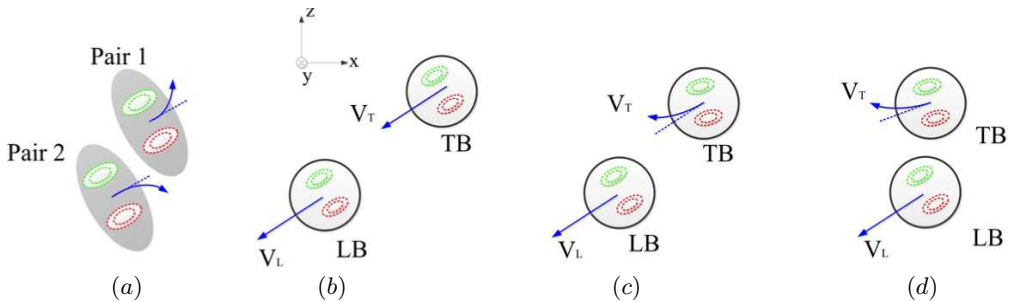}}
  \hspace{15mm}$(a)$ \hspace{22mm}$(b)$\hspace{35mm}$(c)$\hspace{35mm}$(d)$
  \caption{Sketch illustrating the mechanism driving the horizontal precession of the TB path observed for $(Ga,Bo) =(90,0.5$) (bottom view). $(a)$: deviation of a vortex pair in the presence of another vortex pair (the red and green dashed lines schematize the positive and negative vortex threads); $(b)-(d)$: successive stages of the clockwise deviation of the TB as it approaches the LB and undergoes the influence of its wake. The blue arrows indicate the direction of the horizontal drift of the corresponding bubble/vortex pair.}
\label{f3.4.4}
\vspace{-2mm}
\end{figure}
The reason why the interaction process described above results in a precession of the TB is easily understood using figure \ref{f3.4.4} and fundamental laws of vortex dynamics. 
In this figure, both bubbles are drifting toward the left and the streamwise vorticity is negative (resp. positive) in the upper (resp. lower) thread of both pairs (panel $(a)$). This corresponds to an antagonistic interaction configuration in which the induced velocity field tends to make the two pairs repel each other. 
\color{black} For this reason, the vortex pair 1 (corresponding to the TB) is  deviated in the anticlockwise direction, making the TB rotate clockwise since the force acting on the bubble is opposite to that acting on the fluid (panels $(b)-(d)$). The same reasoning indicates that the LB rotates anticlockwise. 
\color{black} 


\begin{figure}
 \vspace{6mm}
  \centerline{\includegraphics[width=1.02\textwidth]{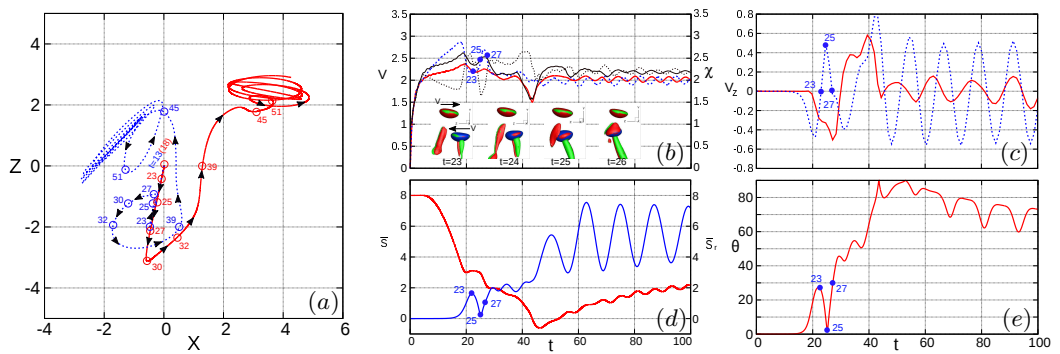}}
  \caption{Evolution of several characteristics of a bubble pair with $(Ga,Bo) =(50,0.5)$. $(a)$: bottom view of the path of the LB (solid line) and TB (dashed line); 
 $(b)$: rising speed (colored lines, left axis) and aspect ratio (black lines, right axis); $(c)$: Z-component of the horizontal velocity; $(d)$: vertical (red, left axis) and horizontal (blue, right axis) separations; $(e)$: inclination of the line of centres. 
 The embedded snapshots in $(b)$ show the iso-surfaces $\omega_y = \pm 3$ past the two bubbles during the time interval $23 \leq t \leq 26$.
 }
 \vspace{-51.5mm}
 \hspace{84mm}$(b)$ \hspace{40.5mm}$(c)$\\
  \vspace{12.5mm}
  
  \hspace{38mm}$(a)$ 
    \vspace{-1.3mm}
 
 \hspace{84mm}$(d)$ \hspace{40.5mm}$(e)$\\
   \vspace{22mm} 
\label{f3.4.5}
\end{figure}
Another example of interacting paths is displayed in figure~\ref{f3.4.5}, the control parameters being now $(Ga, Bo) = (50, 0.5)$.
Similar to the previous case, the two bubbles start to drift in the same direction, with the TB well ahead of the LB (panel $(b)$), implying little influence of the wake of the latter on the TB in this early stage. 
Then at $t = 23$, the horizontal motion of the TB changes sign, which somewhat later ($t = 25$) brings the two bubbles back to the in-line configuration, 
implying $\overline{S}_r \approx 0$ and $\theta \approx 0$ (panels $(d)$ and $(e)$). During this second stage, the dynamics of the TB is strongly influenced by the LB wake, 
the streamwise vortices shed downstream hitting directly the TB as the iso-contours of the vertical vorticity in figure \ref{f3.4.5}$(b)$ indicate. This intense interaction first manifests itself in the sharp decrease (resp. increase) of the TB aspect ratio (resp. rising speed), as the dashed curves in the same panel confirm. 
Then a violent deflection of the TB path toward the $X<0$ direction takes place at $t = 27$, initiating a spiralling anticlockwise motion. This is also a consequence of the above interaction: as a slight misalignment of the two vertical planes in which the paths take place has developed gradually (see the successive positions of the blue and red circles beyond $t=23$ in panel ($a$)),  the impulse transferred to the TB by the streamwise vortices has a nonzero transverse component with respect to its ongoing path, from which the observed lateral deflection follows. The reaction of the LB takes place somewhat later ($t=30$), resulting in a similar and even more abrupt deflection of its path in the opposite direction ($X>0$). \\
\indent This intense sequence is succeeded by a long transitional stage ($30\lesssim t \lesssim45$) during which the two bubbles follow somewhat erratic paths and the intensity of their interaction decays, owing to the gradual increase of their horizontal separation. At $t=45$, the tandem is close to the side-by-side configuration (panel $(e)$). Then, each path evolves almost independently in a much more conventional way. On the one hand, the TB follows a large-amplitude planar zigzagging path with a crest-to-crest amplitude of $3-4$ bubble radii. On the other hand, the LB describes a flattened spiral with a major axis $2-2.5$ bubble radii long. Interestingly, the bubble rotation switches from anti-clockwise to clockwise at some point. The lateral excursions of the TB being larger than those of the LB, the horizontal velocities follow the same trend. This leaves a smaller fraction of the potential energy available for the rise of the TB, making the rising speed of the LB slightly larger, which in turn tends to increase the vertical separation  as panel $(d)$ confirms. In summary, the dynamics examined here represents an intermediate case. The two bubbles strongly interact for $t\lesssim40$, a first `life' during which the system stands in the IFS regime. Then, they evolve almost independently with the average direction of their oscillations lying in two different vertical planes. This makes their behaviour similar to that observed in the NIZ regime, except that here the LB follows a very flattened spiralling path instead of a strictly planar zigzagging path. 

\section{Influence of some parameters}\label{sec4}

\subsection{Influence of the Bond number}\label{sec4.1}

\begin{figure}
 \vspace{6mm}
  \centerline{\includegraphics[width=0.65\textwidth]{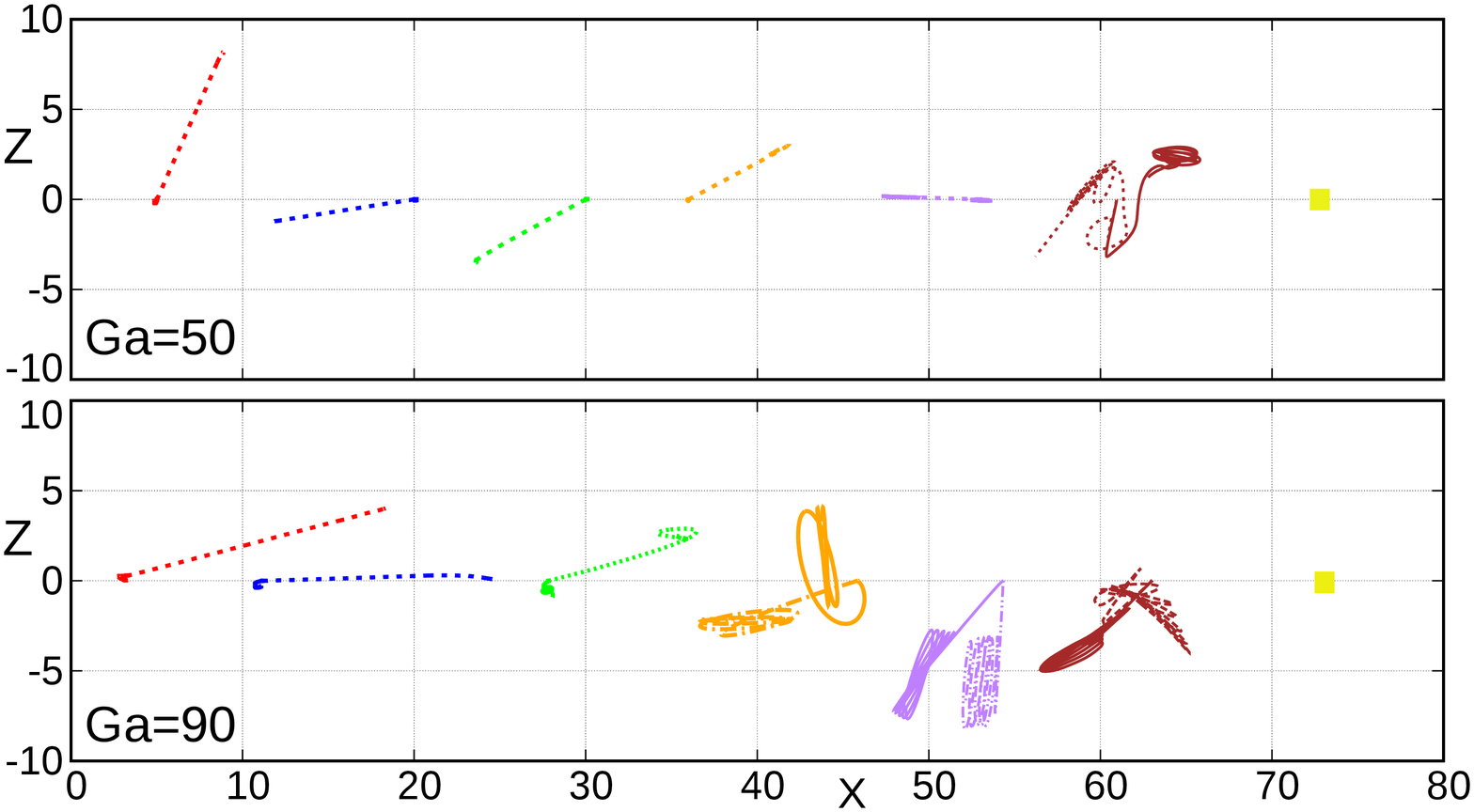}}
  \caption{Bottom views of the paths for various $Bo$ at $Ga = 50$ and $Ga = 90$. From left to right: $Bo = 0.02, 0.05, 0.1, 0.2, 0.3, 0.5$ and $1.0$ (the latter with a yellow square corresponding to a head-on collision). Solid and dashed lines refer to the LB and TB, respectively. A shift has been applied to the initial $X$-position of each pair for readability. }
\vspace{1mm}
\label{f4.1.1}
\end{figure}


The influence of the Bond number on the fate of the bubble pair was discussed in detail in ZNM for $Ga\leq30$. It was shown that the direct connection between the bubble oblateness and the magnitude of the vorticity generated at the gas-liquid interface results in a strong increase of the wake `sheltering' effect with the Bond number. For this reason, the larger $Bo$ is, the stronger the attraction of the TB toward the LB is and the later its lateral escape starts. This is what makes the two bubbles eventually collide and coalesce beyond the critical Bond number $Bo_c$ introduced earlier. A side effect is that, for $Bo<Bo_c$, the final horizontal separation of the two bubbles decreases as $Bo$ increases, given the shorter time during which the TB is able to drift. The above mechanism still holds in the parameter range considered here. However, in this higher-$Ga$ range, the fact that vortex shedding takes place  makes the horizontal separation reached after the  initial drift of the TB a crucial parameter for the possibility of further interactions. In other words, increasing $Bo$ (still with $Bo<Bo_c$) makes the system shift from the CIZ or NIZ regimes to the IFS regime. \\
\indent Figure~\ref{f4.1.1}, which displays the horizontal trace of the paths, confirms the expected tendency. For $Ga = 50$, the gradual reduction of the horizontal separation as $Bo$ increases is clearly seen in the ASE regime ($Bo\leq0.2$). Then, the dramatic change of the trace from $Bo=0.3$ to $0.5$ helps appreciate how the mechanisms involved in the interaction of the two bubbles modify the geometry of their paths during the succession of the CIZ, NIZ and IFS regimes. The change of the horizontal trace through the NIZ-IFS transition that takes place in the range $0.4<Bo<0.5$ is also well visible for $Ga=90$. At this high $Ga$, bubble pairs with $Bo\lesssim0.1$ rise in the SIE regime. This is why in figure \ref{f4.1.1} the two bubbles of the corresponding pairs are seen to perform only tiny horizontal displacements once the initial ASE drift is completed. 

\subsection{Influence of an initial angular deviation}\label{sec4.2}
\begin{figure}
  \vspace{6mm}
  \centerline{\includegraphics[width=0.5\textwidth]{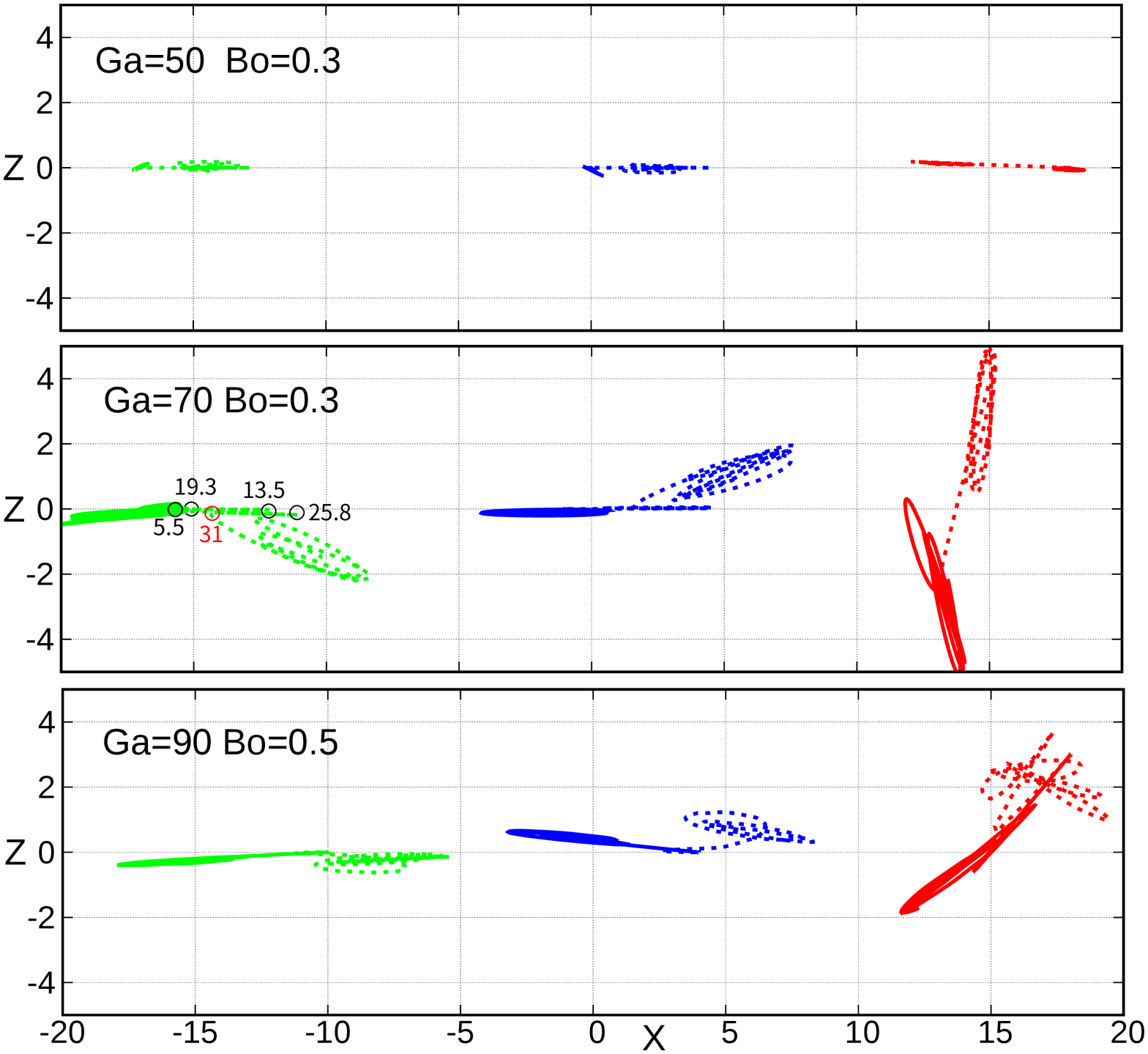}}
  \caption{Variations of the path geometry (bottom view) with the initial angular deviation $\theta_0$ for three typical bubble pairs which, in the reference case $\theta_0=0^\circ$, evolve in the CIZ, NIZ and IFS regimes (from top to bottom). 
  Red, blue and green lines correspond to $\theta_0=0^{\circ}$, $1^{\circ}$ and $2^{\circ}$, respectively. A shift in the initial $X-$position has been applied to each pair for readability; the numbers along the green dashed line in the central row indicate the corresponding time instant. Initially, the two bubbles stand a dimensionless horizontal distance $\overline{S}_r=\overline{S}_0\tan\theta_0$ apart, with the TB shifted in the $X>0$ direction. }
  \vspace{0mm}
\label{f4.2.1}
\end{figure}
In ZNM it was shown that for $Ga\leq30$ a small nonzero initial inclination $\theta_0$ between the two bubbles has a dramatic influence on their fate. The initial configuration being already non-axisymmetric in this case, the lateral drift of the TB starts much earlier, i.e. at significantly larger separations, than in the $\theta_0=0^\circ$ case. As a consequence, the DKT interaction which characterizes the dynamics for $Ga=\mathcal{O}(10)$ and $Bo\lesssim0.2$ when $\theta_0=0^\circ$ no longer exists with an initial inclination as weak as $2^\circ$. The critical Bond number beyond which the two bubbles collide and eventually coalesce also increases significantly with $\theta_0$, owing to the longer time offered to the TB to leave the LB wake. Last,  
when the two bubbles do not collide, their final horizontal separation is smaller than in the $\theta_0=0^\circ$ case, since the shear that drives the lateral drift of the TB is weaker, owing to the larger $\overline{S}$ at which this drift takes place. Here we question the influence of $\theta_0$ on the dynamics of the bubble pair at higher $Ga$. As an introduction, figure~\ref{f4.2.1} shows how the horizontal trace of the paths varies when the initial inclination is increased up to $2^\circ$. Clearly this parameter has a dramatic influence on the three-dimensionality of the path, tending to constrain it to develop in (or close to) the vertical plane containing initially the two bubbles, here the $(X,Y)$ plane. The reason for this is clear. For an isolated bubble, path instability arises through a bifurcation that preserves a symmetry plane in the wake \citep{mougin2001path,Tchoufag2014}. The orientation of this plane is arbitrary and is usually dictated by some initial disturbance. Here, due to the nonzero $\theta_0$, the flow past each bubble does not strictly preserve an axial symmetry about the vertical direction, even at short time. For this reason, the bifurcation leading to non-straight paths is imperfect, the $(X,Y)$ plane providing a preferential orientation to the symmetry plane. Therefore, the zigzagging motion of the TB resulting from this imperfect bifurcation takes place in this preferential plane, until the LB wake possibly provides a subsequent disturbance having a component out of that plane. Figure~\ref{f4.2.1} confirms that this scenario holds in all cases. Because of the presence of a preferential vertical plane, configurations such as that found for $Ga=90,Bo=0.5$ with $\theta_0=0^\circ$, for which the two paths evolve in a complicated way and eventually stabilize in nearly perpendicular planes, no longer exist. Instead, for the same $Ga$ and $Bo$ but $\theta_0=2^\circ$, the influence of the initial angular deviation is seen to be sufficient to make both bubbles perform large-amplitude nearly planar zigzags within vertical planes whose orientation almost coincides with that of the $(X,Y)$ plane. A major consequence of this modified geometry is that, beyond the initial ASE stage, the two bubbles maintain a sufficient separation for their interaction to be very weak. In other terms, while the bubble pair with $Ga=90,Bo=0.5$ belongs to the IFS regime when $\theta_0=0^\circ$, it rather stands in the CIZ regime when $\theta_0=2^\circ$. However, as we shall see below, this is not a general rule, even though the two paths take place within the same vertical plane. \\ 
\begin{figure}
\vspace{6mm}
  \centerline{\includegraphics[width=0.98\textwidth]{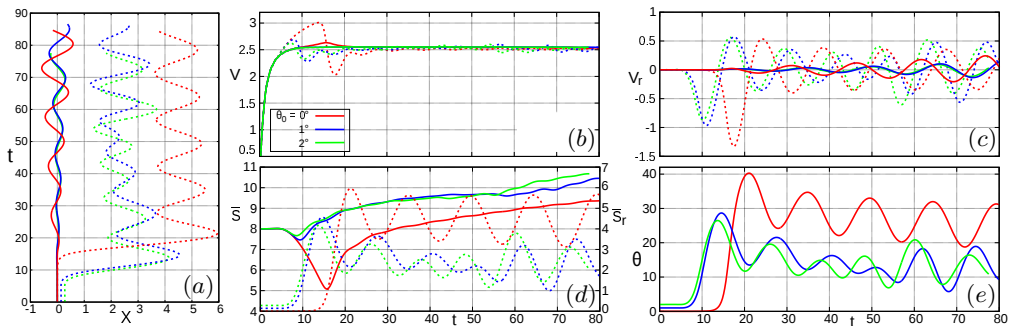}}
  \caption{Evolution of several characteristics of a bubble pair with $(Ga,Bo) = (50,0.3)$ undergoing initial angular deviations $\theta_0$ from $0^{\circ}$ to $2^{\circ}$. $(a)$: side view of the LB (solid line) and TB (dashed line) paths; $(b)$: rising speed; $(c)$: transverse velocity; $(d)$: vertical (solid lines, left axis) and transverse (dashed lines, right axis) separations; $(e)$: inclination of the line of centres.}
   \vspace{-43mm}
 \hspace{75.3mm}$(b)$ \hspace{48mm}$(c)$\\
  \vspace{12.3mm}
  
  \hspace{24.5mm}$(a)$ 
    \vspace{-2.3mm}
 
 \hspace{75.1mm}$(d)$ \hspace{47.7mm}$(e)$\\
   \vspace{12mm} 
\label{f4.2.2}
\end{figure}
\indent To illustrate the influence of the initial deviation in more detail, we consider the bubble pair with $(Ga, Bo) = (50, 0.3)$. As figure \ref{f4.2.1} revealed, any nonzero initial deviation makes the corresponding tandem rise within the $(X,Y)$ plane. The earlier start of the TB drift triggered by the slightly asymmetric initial configuration is clearly seen in figures \ref{f4.2.2}$(a)$ and $(c)$. This earlier start has a direct consequence on the future rise of the LB: since $\overline{S}$ is larger than in the reference case during the TB drift ($\overline{S}\approx8$ instead of $\overline{S}\approx5-6$ for $\theta_0=0^\circ$ according to figure \ref{f4.2.2}$(d)$), the disturbance induced near the LB is smaller, delaying the triggering of its path instability and reducing its growth rate (compare the blue and red solid curves in panels $(a)$ and $(c)$). The weaker shear rates encountered at this larger $\overline{S}$ by the TB during its drift across the LB wake drastically reduce the lateral extension of the paths in the $X$-direction, i.e. the maxima of $\overline{S}_r$, confirming the indication of figure \ref{f4.2.1}. Indeed, the mean value of $\overline{S}_r$ which is close to $4.5$ when $\theta_0=0^\circ$ is reduced to $2.1$ for both $\theta_0=1^\circ$ and $\theta_0=2^\circ$ (figure \ref{f4.2.2}$(d)$). The large-amplitude zigzags performed by the TB and their phase difference with the slowly growing oscillations of the LB make $\overline{S}_r$ experience large variations, with crest-to-crest amplitudes as large as $2.5$ in the late stage of the simulations. The fact that the minima of the horizontal separation periodically reach values in the range $1\lesssim\overline{S}_r\lesssim2$ while the vertical separation is large ($\overline{S}\gtrsim9$) implies that the TB repeatedly comes back into the LB wake. Because of this, the dynamics of the TB remain influenced by its interaction with that wake throughout its rise. Actually, this interaction even tends to strengthen as time proceeds, the vertical separation increasing by only $35\%$ from $t=20$ to $t=80$ while the minimal inclination of the tandem reduces by nearly $60\%$ and becomes less than $10^\circ$ for $t>50$ (figure \ref{f4.2.2}$(e)$). This continuous interaction, combined with the large amplitude of the zigzags performed by the TB, makes $V_{TB}$ experience growing periodic variations with a relative magnitude up to $5\%$ in the late stage; in contrast the rising speed of the LB stays remarkably constant (and independent of $\theta_0$) beyond $t=20$ (figure \ref{f4.2.2}$(b)$). In summary, the evolution of this bubble pair for $\theta_0\neq0^\circ$ reveals the existence of a regime that was not encountered in the reference case: the system still evolves within a single vertical plane, both bubbles perform large-amplitude zigzagging motions, but the TB continues to experience the influence of the LB wake throughout its rise, far from the `independent' evolution discussed in \S\,\ref{sec3.2}. 
\\
\begin{figure}
\vspace{6mm}
  \centerline{\includegraphics[width=0.98\textwidth]{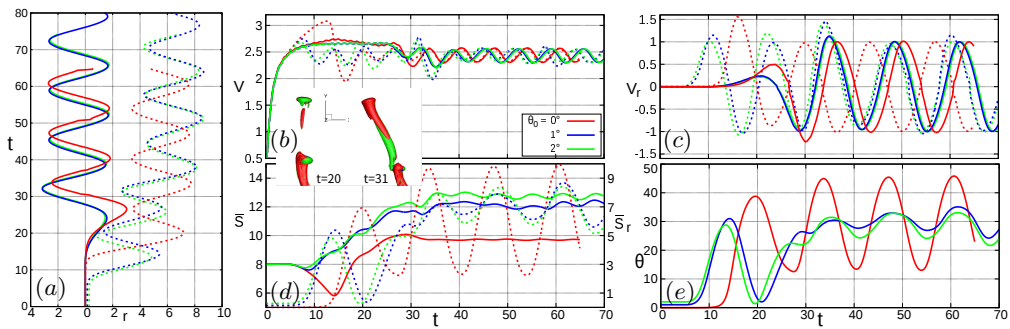}}
  \caption{Evolution of several characteristics of a bubble pair with $(Ga,Bo) = (70,0.3)$ undergoing initial angular deviations $\theta_0$ from $0^{\circ}$ to $2^{\circ}$. For caption, see figure~\ref{f4.2.2}. The two snapshots straddling panels $(b)$ and $(d)$ show the iso-surfaces $\omega_y = \pm 0.5$ at two different instants of time for $\theta_0=2^\circ$.}
  \vspace{-43mm}
 \hspace{36mm}$(b)$ \hspace{47mm}$(c)$\\
  \vspace{12.5mm}
  
  \hspace{5mm}$(a)$ 
    \vspace{-3mm}
 
 \hspace{36mm}$(d)$ \hspace{47mm}$(e)$\\
   \vspace{14mm} 
\label{f4.2.3}
\end{figure}\indent
The influence of $\theta_0$ on the evolution of the bubble pair discussed in \S\,\ref{sec3.3} $(Ga=70, Bo=0.3)$ is also worthy of some comments. The characteristics of the corresponding dynamics are detailed in figure~\ref{f4.2.3}, the bottom view of the paths being displayed in the central row of figure \ref{f4.2.1}. The general trends of this dynamics are similar to those already identified in the strictly in-line configuration. In short, the two bubbles soon develop large-amplitude planar zigzagging motions and, beyond a certain stage, evolve independently from each other in two distinct vertical planes. This makes this pair belong to the NIZ regime whatever $\theta_0$ in the range $0^\circ-2^\circ$. What is of most interest here is the way the two bubbles interact before this `independent' state, especially the processes that lead to the selection of the vertical plane in which each of them eventually rises. In the original in-line configuration, the vertical plane in which the LB oscillates is selected by the instability that develops in its near wake (as in the case of an isolated bubble with similar characteristics), while the TB oscillates in a plane whose orientation is dictated by its initial drift. Since the two selection mechanisms are independent, the two planes have different orientations from the very beginning of the LB oscillations. This is no longer the case for $\theta_0=1^\circ$ and $2^\circ$. Here, not only is the flow asymmetry introduced by the initial inclination of the bubble pair sufficient to select the direction of the initial TB drift, but it also almost controls the orientation of the symmetry plane resulting from the (now imperfect) bifurcation initiating the zigzagging motion of the LB. Therefore the TB oscillates within the $(X,Y)$ plane and the LB does the same within a plane making only a small angle with the former. Obviously, this configuration makes the TB more prone to experience disturbances that later emanate from the LB wake. This could already be the case at $t\approx20$: at this instant, the TB has already completed its first zigzag while the LB only starts its own, so that the large phase difference makes the two bubbles almost realign vertically (figures \ref{f4.2.3}$(a)$, $(d)-(e)$). However, as the first snapshot inserted in panels $(b)$ and $(d)$ reveals, the streamwise vortices in the LB wake are still weak and the vertical separation is large ($\overline{S}\approx10$). Therefore the TB is barely disturbed by the LB wake and starts a new zigzag, still in the $(X,Y)$ plane. Then, the oscillations of the LB path saturate quickly ($t\approx25$), so that the streamwise vortices soon develop downstream and reach the TB at $t\approx31$ (second snapshot in panels $(b)$ and $(d)$), although the two bubbles are widely separated at this stage ($\overline{S}\approx12,\,\overline{S}_r\approx5.5$). Since the symmetry plane of the LB wake makes a small but nonzero angle with the $(X,Y)$ plane, the momentum transferred to the TB during the collision with the streamwise vortices has a nonzero component perpendicular to the latter plane. This deflects the path of the TB toward another plane distinct from the previous two. From there on, the possibility for another interaction sequence of the same nature is much reduced, and the system now evolves in a NIZ configuration very similar to that observed for $\theta_0=0^\circ$.   

\subsection{Influence of the initial separation in the zigzagging/spiralling regimes}\label{sec4.3}

\begin{figure}
\vspace{6mm}
  \centerline{\includegraphics[width=0.7\textwidth]{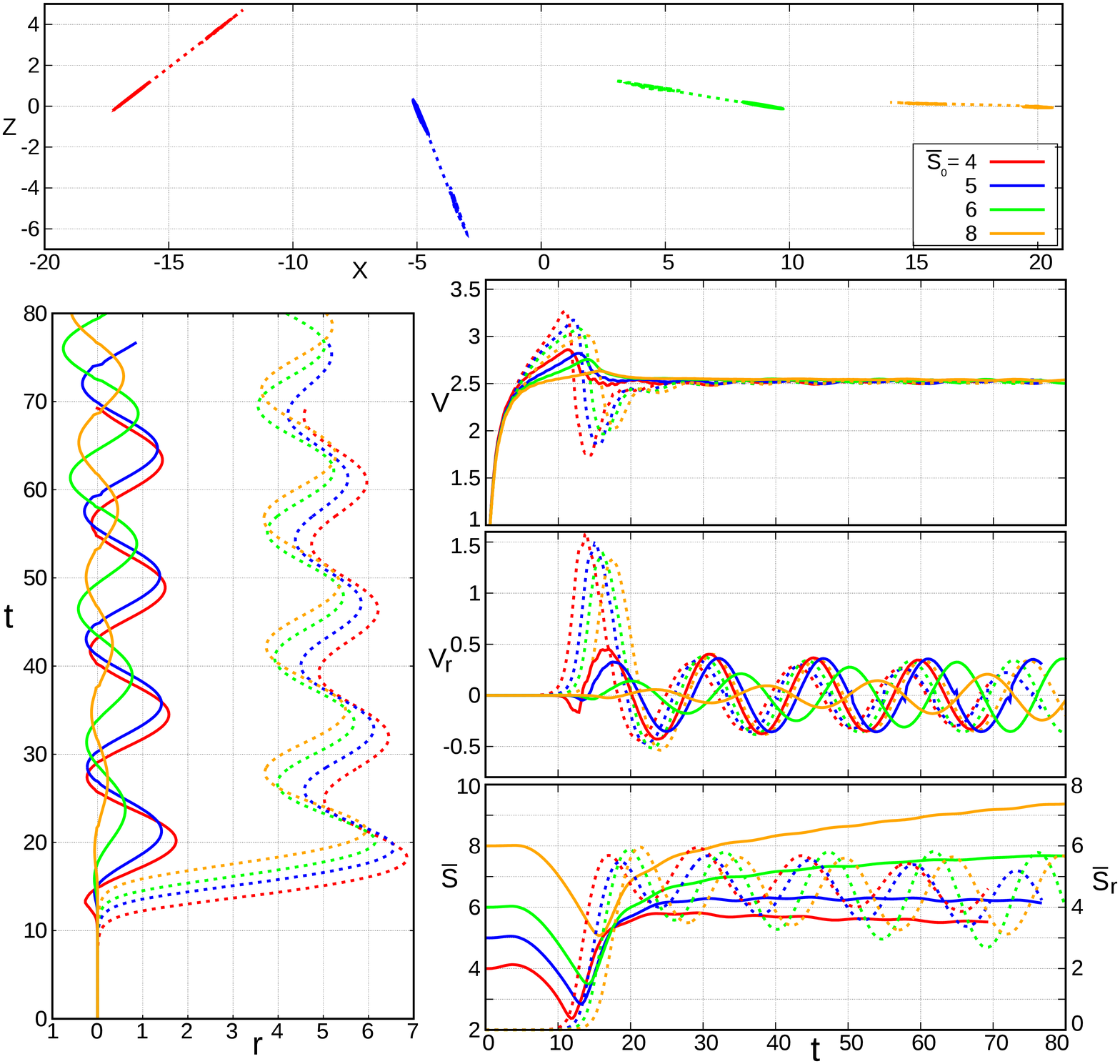}}
  \caption{Evolution of several characteristics of a bubble pair with $(Ga,Bo) = (50,0.3)$ for initial separations in the range $4\leq\overline{S}_0\leq8$. $(a)$ and $(b)$: bottom and side views of the path of the LB (solid line) and TB (dashed line), respectively; $(c)$ and $(d)$: rising speed and transverse velocity of the two bubbles (same convention), respectively; $(e)$ vertical (solid line, left axis) and transverse (dashed line, right axis) separations.}
  \vspace{-92mm}
  \hspace{17mm}$(a)$\\
  
  \vspace{15mm}\hspace{62mm}$(c)$\\
  
   \vspace{14.5mm}\hspace{62mm}$(d)$
 
    \vspace{17.7mm}\hspace{17mm}$(b)$\hspace{41mm}$(e)$
   \vspace{21mm}
\label{f4.3.1}
\end{figure}

So far the initial separation between the two bubbles was maintained at $\overline{S}_0 = 8$. We now examine how varying $\overline{S}_0$ may affect the interaction process in the parameter range where the CIZ and NIZ scenarios were previously encountered. 
Figure~\ref{f4.3.1} displays several characteristics of the dynamics obtained by decreasing $\overline{S}_0$ from $8$ to $4$ for $(Ga,Bo)=(50,0.3)$. 
As the bottom view of the paths in panel $(a)$ indicates, the system keeps its planar symmetry and the average horizontal separation does not change much as $\overline{S}_0$ is varied (see panels $(b)$ and $(e)$ for a quantitative confirmation). Therefore, the bubble pair remains in the CIZ regime at least down to $\overline{S}_0=4$. The minimum reached by the vertical separation during the initial axisymmetric stage decreases significantly as $\overline{S}_0$ is reduced, from $\overline{S}\approx5$ for $\overline{S}_0 = 8$ to $\overline{S}\approx2.3$ for $\overline{S}_0 = 4$. Moreover, the smaller $\overline{S}_0$ the shorter the time required to reach this minimum. A direct consequence of this shorter initial stage, already observed at lower $Ga$ in ZNM, is that the smaller $\overline{S}_0$ is, the earlier the initial lateral drift of the TB starts, as panels $(b)$ and $(d)$ confirm. At the same time, the reduction of the $\overline{S}$-minimum implies that the magnitude of the asymmetric disturbance induced by this drift in the vicinity of the LB increases as  $\overline{S}_0$ is reduced, triggering the lateral motion of the latter. This is why the zigzagging motion of the LB reaches a saturated state much earlier for $\overline{S}_0\leq5$ ($t\approx20$) than for larger initial separations ($t \gtrsim 120$ for $\overline{S}_0 = 8$, see figure~\ref{f3.2.2}). The behaviour of the system for $\overline{S}_0>8$ may be extrapolated by combining the information provided by figure~\ref{f4.3.1}$(e)$. First, since the $\overline{S}$-minimum reached during the initial stage increases with $\overline{S}_0$, the two bubbles cannot collide during that stage, and the initial drift of the TB out of the LB wake is likely to take place whatever $\overline{S}_0$. Beyond this stage, the long-term value of $\overline{S}$ increases monotonically with $\overline{S}_0$ while the average lateral separation stays almost $\overline{S}_0$-independent. From this monotonic behaviour in the range $4\leq\overline{S}_0\leq8$, it may reasonably be inferred that the system remains in the CIZ regime for $\overline{S}>8$, with possible significant changes only in the final vertical separation, hence in the inclination of the tandem. A qualitatively similar conclusion is reached by considering the pair with $(Ga,Bo)=(70,0.3)$ and varying the initial separation in the range $5\leq\overline{S}_0\leq20$ (not shown): the system stays in the NIZ regime whatever $\overline{S}_0$, although its final geometry, especially the vertical separation of the two bubbles and the orientation of their line of centres, varies significantly with the initial separation.

\begin{figure}
\vspace{6mm}
  \centerline{\includegraphics[width=0.98\textwidth]{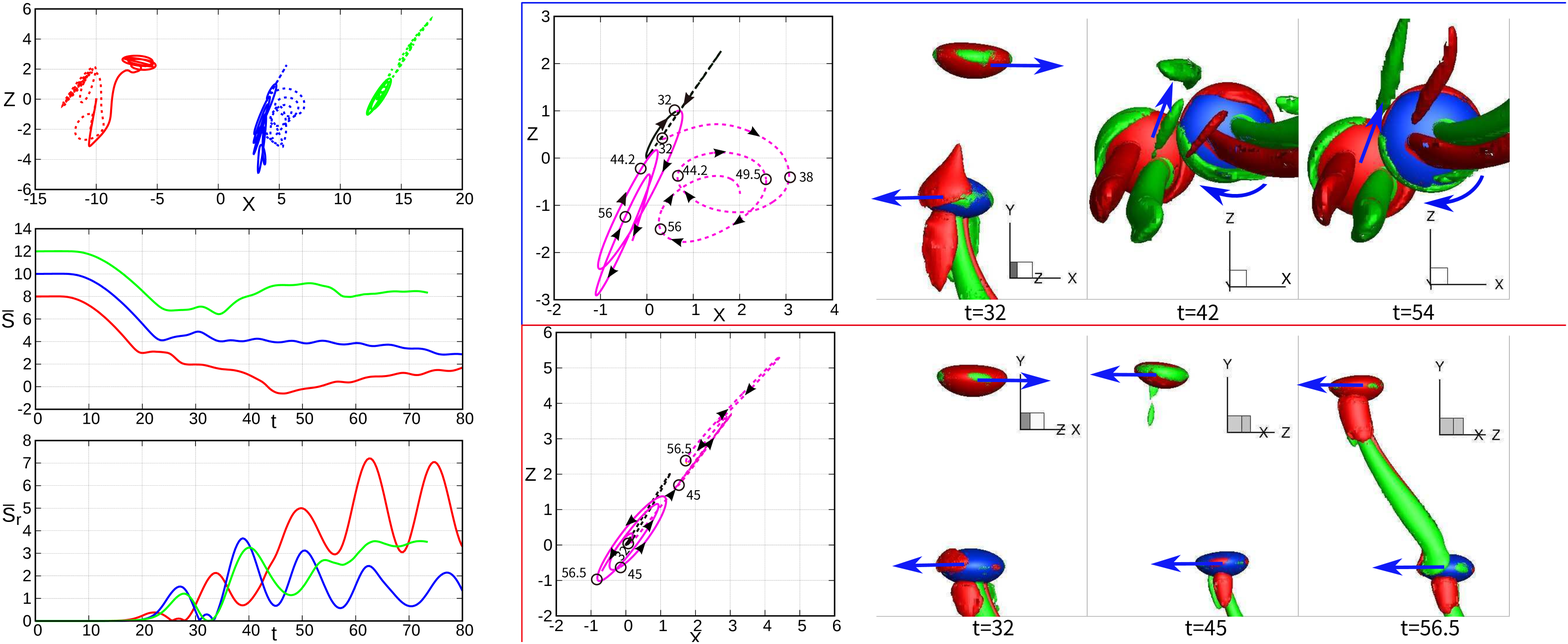}}
  \caption{Evolution of several characteristics of a bubble pair with $(Ga,Bo) = (50,0.5)$ for initial separations in the range $8\leq\overline{S}_0 \leq12$. $(a)$: bottom view of the paths; $(b)-(c)$: vertical and transverse separations, respectively; $(d)-(e)$: enlarged horizontal trace of the paths (left panel) and vertical vorticity distribution (identified with the iso-surfaces $\omega_y=\pm1$) at different instants of time in various vertical or horizontal planes for $\overline{S}_0 = 10$ and $\overline{S}_0 = 12$, respectively. In $(a)-(c)$ the red, blue and green lines correspond to $\overline{S}_0 =8$, $10$ and $12$, respectively. In the insets of $(d)-(e)$, the black (resp. purple) part of the paths corresponds to the rectilinear (resp. spiralling/zigzagging) stage that precedes (resp. succeeds) the realignment of the two bubbles at $t\approx32$.} 
  \vspace{-77.2mm}\hspace{35.5mm}$(a)$\\
  
  \vspace{4.6mm}\hspace{129mm}$(d)$\\
  
  \vspace{-0.1mm}\hspace{35.8mm}$(b)$\\
  
   \vspace{10.9mm}\hspace{35.7mm}$(c)$\\
   
    \vspace{-7mm}\hspace{129mm}$(e)$\\
  
   \vspace{28mm}\
\label{f4.3.2}
\end{figure}

We then examine the changes in the dynamics of the pair with $(Ga, Bo) = (50, 0.5)$ when the initial separation is varied from $\overline{S}_0 = 8$ to $\overline{S}_0 = 12$. 
 Figure~\ref{f4.3.2}$(a)$ indicates that increasing the initial separation makes the horizontal projection of the paths flatter. Qualitatively, the system first transitions from the mixed IFS-NIZ evolution described in \S\,\ref{sec3.4} for the reference case $\overline{S}_0 = 8$ to a clear IFS evolution for $\overline{S}_0 = 10$. Then, for $\overline{S}_0 = 12$, the two paths are almost coplanar, so that the observed dynamics look like a CIZ evolution, with the difference that the LB follows a very flattened spiralling path rather than a strictly planar zigzagging motion. For both  $\overline{S}_0 = 10$ and  $\overline{S}_0 = 12$, the tandem first exhibits an evolution qualitatively similar to that described in \S\,\ref{sec3.4}, with essentially some time shift. For instance, the system realigns at $t\approx31-32$, instead of $t\approx25$ in the reference case. This realignment leads potentially to a direct interaction between the TB and the trailing vortices shed by the LB. However, this interaction weakens dramatically as $\overline{S}_0$ increases as we shall see. This is because the LB stands right at the threshold of path instability, its aspect ratio being very close to $2.2$. Under such nearly critical conditions, the disturbance provided by the initial drift of the TB is crucial for triggering the path instability of the LB (see the discussion in \S\,\ref{sec3.3}). Therefore, the larger $\overline{S}_0$ is, the longer it takes for the trailing vortices in the LB wake to grow.\\
 \indent For $\overline{S}_0 = 10$, the LB wake is significantly but not completely developed by the time the two bubbles realign. Hence, similar to the reference case (figure \ref{f3.4.5}), the TB undergoes a violent lateral deviation after the tandem has realigned vertically at $t\approx31$. However, here the two bubbles start to spiral right away, while for $\overline{S}_0 = 8$ they drift erratically during a long transient before starting to spiral. The absence of this transient is a consequence of the weaker lateral momentum transferred by the trailing vortices to the TB in the present case, owing to the aforementioned incomplete development of the LB wake and the larger vertical separation at the corresponding time ($\overline{S}\approx4.5$ instead of $\overline{S}\approx2$, according to figure~\ref{f4.3.2}$(b)$). As a result, the horizontal separation between the two bubbles stays much smaller than in the reference case, even at large time (see figure \ref{f4.3.2}$(c)$, where the $\overline{S}_r$-minima are seen to be now about $\overline{S}_r\approx0.7$, instead of $\overline{S}_r\approx3$ for $\overline{S}_0 = 8$). Hence, as the last two snapshots in figure \ref{f4.3.2}$(d)$ confirm, the TB keeps on being under the influence of the LB wake at regular time intervals throughout its rise, providing a new example of an IFS evolution. \\
 \indent Beyond $t\approx32$, the evolution observed when $\overline{S}_0 = 12$ differs dramatically from the above one. Indeed, as the orientation of the black and purple segments in the left panel of figure \ref{f4.3.2}$(e)$ indicate, the TB deviates only by a small angle from the path it followed in the earlier stage. For the aforementioned reason, the strength of the trailing vortices past the LB is still weak in this case by the time the two bubbles realign (left snapshot in figure~\ref{f4.3.2}$(e)$). Moreover, the vertical separation is significantly larger than with $\overline{S}_0 = 10$ ($\overline{S}\approx7$ instead of $\overline{S}\approx4.5$). Both features cooperate to limit the influence of the LB wake on the TB during this crucial stage, so that the initial planar symmetry of the system is only marginally altered. At longer times, the TB essentially performs large planar zigzags. However it may occasionally interact with the LB wake which is now fully developed. During such interactions, some lateral momentum is transferred to the TB, inducing some slow precession of its plane of oscillation (see the right snapshot in figure ~\ref{f4.3.2}$(e)$ and the change in the horizontal trace at $t=56.5$).

\subsection{Influence of the initial separation in the collision regime}\label{sec4.4}

\begin{figure}
\vspace{6mm}
  \centerline{\includegraphics[width=0.98\textwidth]{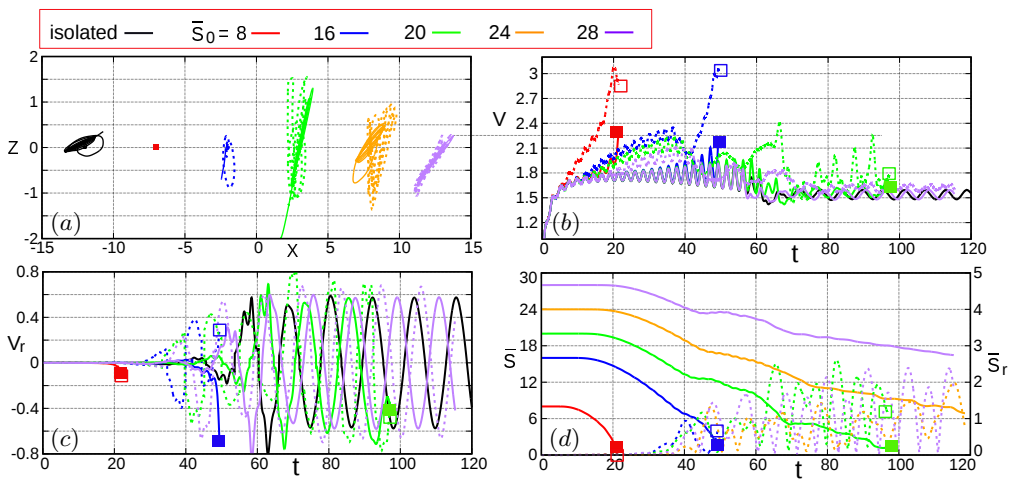}}
  \caption{Evolution of several characteristics of a bubble pair with $(Ga,Bo) = (50,1.0)$, with the initial separation varied in the range $8\leq\overline{S}_0\leq28$. $(a)$: bottom view of the paths; $(b)-(c)$: vertical and transverse velocities of the LB (solid line) and TB (dashed line), respectively; $(d)$: vertical (solid line, left axis) and transverse (dashed line, right axis) separations. The squares in $(b)-(d)$ indicate the point at which the two bubbles collide. In $(a)$ a shift in the initial $X$-position has been applied to each pair for readability, while in $(b)-(c)$ the results corresponding to $\overline{S}_0=24$ are not shown for the same reason.}
   \vspace{-63.5mm}\hspace{7mm}$(a)$\hspace{62mm}$(b)$\\
    
   \vspace{21.3mm}\hspace{7mm}$(c)$\hspace{62mm}$(d)$\\
  
   \vspace{28mm}

\label{f4.4.1}
\end{figure}

Here we consider the parameter set $(Ga, Bo) = (50,1.0)$ and vary the initial separation to analyze its influence in the collision regime identified for $\overline{S}_0=8$ in figure \ref{f3.1.1}. 
Some characteristics of the dynamics observed by increasing the initial separation up to $\overline{S}_0=28$ are displayed in figure~\ref{f4.4.1}. The two bubbles are found to collide for $\overline{S}_0 \leq 20$, the collision time increasing significantly with $\overline{S}_0$ (figures~\ref{f4.4.1}$(b)-(d)$). They stay significantly apart from each other until the end of the simulation for the largest two separations. Nevertheless, as figure \ref{f4.4.1}$(d)$ reveals, the vertical separation decreases consistently over time whatever $\overline{S}_0$, suggesting that the two bubbles would also eventually collide later for these large initial separations. Present observations are reminiscent of those of \cite{Stewart1995} and \cite{brucker1999structure} with large rising bubbles having Bond numbers in the range $1-10$ and  Galilei numbers in the range $100-500$. In these experiments, the zigzagging/spiralling bubbles were always found to collide despite the lateral excursions of the TB. Coalescence never happened, presumably because the interfaces were contaminated in these experiments. Instead, the bubbles repelled each other after collision, leading to a self-repeating scenario.\\
\indent According to figure~\ref{f4.4.1}$(a)$, an isolated bubble with the same $Ga$ and $Bo$ (black line) first follows a flattened spiralling path before describing planar zigzags, an evolution consistent with the findings of CL16. Nevertheless, it takes some time for the initial straight vertical path to become unstable, which explains why the two bubbles collide in the head-on configuration for $\overline{S}_0 = 8$. Path instability of the isolated bubble sets in at $t\approx25$ and saturates at $t\approx70$ according to figure~\ref{f4.4.1}$(c)$. Consistently, all pairs with $\overline{S}_0 > 8$ perform oscillatory paths. For $\overline{S}_0 > 16$, the saturated path of the LB takes the form of a planar zigzag or (for $\overline{S}_0 =24$) a flattened spiral. Simultaneously, the TB follows a slowly precessing zigzagging path. A noticeable feature is that the amplitude of the horizontal excursions decreases with $\overline{S}_0$, approximately from $2.5$ for $\overline{S}_0 =20$ to $1.5$ for $\overline{S}_0 =28$, and reaches minima close to zero at regular time intervals (figure \ref{f4.4.1}$(d)$). These characteristics, together with the aforementioned decrease of $\overline{S}$ over time, reinforce the view that a collision happens in all cases. In figure~\ref{f4.4.1}$(b)$, the records of $V_\mathrm{LB}$ are seen to collapse on a single curve, indicating that the LB is barely disturbed by the presence of the TB throughout its rise.  The only exception is the very late stage before a collision happens in the head-on ($\overline{S}_0=8$) or oblique ($\overline{S}_0=16$) configuration, in which case $V_\mathrm{LB}$ sharply increases, owing to the vertical impulse transferred to the LB by the TB which is about to touch it. A  close look at the plots in figures~\ref{f4.4.1}$(b)-(c)$ for $\overline{S}_0 =28$ is of interest. Beyond $t\approx70$, the vertical and transverse velocities of both bubbles appear to have reached a saturated state. However the rising speed of the TB and the maxima of its transverse velocity are slightly larger than those of the LB, indicating that some interaction is still present. Indeed, since the horizontal distance separating the two bubbles remains small, the TB keeps on being sucked toward the LB, due to the slight velocity defect in the far wake of the LB. \\
\indent The key remaining issue is to understand why, as figure \ref{f4.4.1}$(a)$ shows, the average horizontal separation between the two bubbles decreases when $\overline{S}_0$ increases, which leads unavoidably to a collision, and presumably to coalescence in a good number of cases as far as the interfaces are uncontaminated. For this it is necessary to consider the characteristics of the corresponding isolated bubble, whose aspect ratio and terminal Reynolds number are $\chi\approx2.55$ and $Re\approx77$, respectively. \cite{adoua2009reversal} computed the shear-induced lift force acting on an oblate spheroidal bubble held fixed in a linear shear flow over a wide range of conditions. With the above characteristics, their results indicate that, if the relative shear rate (defined as the shear-induced velocity difference at the bubble scale, normalized by the local relative fluid velocity) stands below a $Re$-dependent threshold, the lift force is oriented in the direction opposite to that it would have on a spherical or moderately oblate bubble, other things being equal. The reasons for this change of sign were summarized in appendix C of ZNM. In short, beyond a critical $Re$-dependent oblateness, the sign (but not the strength) of the streamwise vorticity in each of the trailing vortices of the bubble wake is dictated by the vorticity produced at its surface, and no longer by  the upstream vorticity. A detailed analysis of the streamwise vorticity balance reveals that, for a given upstream vorticity, this change results in a switch of the sign of the streamwise vorticity in the trailing vortices, hence in a reversal of the shear-induced lift force acting on the bubble. Quantitatively, this lift reversal may be appreciated by writing the dimensional lift force in the classical form $\mathbf{F}=\mathcal{C}_L\mathcal{M}\mathbf{U}\times\boldsymbol{\omega}_\infty$, with $\mathcal{M}$ the mass of fluid that can be contained within the bubble volume, $\boldsymbol{\omega}_\infty$ the upstream vorticity and $\mathbf{U}$ the relative fluid velocity at the position of the bubble centroid. In the inviscid limit (i.e. with no vorticity generated at the bubble surface by the shear-free condition), the lift coefficient on a spherical bubble is $\mathcal{C}_L=0.5$ \citep{auton1987lift}. Still in this limit, $\mathcal{C}_L$ is an increasing function of the aspect ratio, with $\mathcal{C}_L\approx1.45$ for $\chi=2.55$. These predictions are only marginally altered by finite-$Re$ effects as far the relative shear rate $Sr=R||\boldsymbol{\omega}_\infty||/||\mathbf{U}||$ is such that $Sr\gtrsim0.1$. In contrast, for smaller shear rates, the lift force changes sign beyond $Re\approx37$ for this specific oblateness, and an interpolation of the results of \cite{adoua2009reversal} predicts $\mathcal{C}_L\approx-0.6$ for $Re=77$ and $Sr=0.02$. \\
\indent The manner the above mechanism influences the evolution of the bubble pair in the situation of interest here is as follows. When the TB oscillates along a zigzagging path, the streamwise vorticity in each of its trailing vortices changes sign periodically, as it would for an isolated zigzagging bubble \citep{brucker1999structure,mougin2006wake}. The upstream vorticity $\boldsymbol{\omega}_\infty$ `felt' by the TB results from the axisymmetric velocity defect in the LB wake, and this upstream ($\approx$ horizontal) vorticity is responsible for an additional amount of streamwise ($\approx$ vertical) vorticity in the TB wake, compared to the isolated configuration. When the vertical separation between the two bubbles is moderate, say typically $\overline{S}\lesssim10$, the TB stands in the near wake of the LB. There, the relative shear rate is large enough ($Sr\gtrsim0.1$) for the above two sources of streamwise vorticity to cooperate and provide a shear-induced lift force acting to move the TB away from the wake centreline. Nevertheless, at such short separations, the sheltering effect acting along the wake axis is so strong  that the TB does not succeed in drifting laterally by a sufficient distance to avoid colliding with the LB. In contrast, for larger $\overline{S}$, the above two mechanisms combine in an antagonistic manner, making the streamwise vorticity slightly more intense when the TB is close to the inflection points of the zigzag (i.e. to the centreline of the LB wake) than when it is close to its extremities. This may be confirmed from the curves corresponding to $\overline{S}=20$ and $\overline{S}=28$ in figure \ref{f4.4.1}$(c)$: as the transverse velocity $V_r$ reaches its maxima at the inflection points of the path, the fact that these maxima are slightly larger for the TB than for the LB implies that the lift force responsible for the generation of $V_r$ is somewhat stronger for the former. This unequal magnitude of the lift force limits the amplitude of the lateral excursions of the TB and maintains the midline of its path within the LB wake. The weaker $Sr$, i.e. the larger $\overline{S}$, the more unequal the magnitudes of the lift force acting on the TB during the two halves of a zigzag, hence the shorter the average lateral separation between the two bubbles, as figure \ref{f4.4.1}$(a)$ confirms. 

\section{Summary and concluding remarks}
\label{sec5}
In this second part of our investigation, we carried out three-dimensional numerical simulations of the flow past a pair of identical bubbles initially released in line in the parameter range $40 \leq Ga \leq 90$, $0.02 \leq Bo \leq 1.0$. In that range, provided the Bond number exceeds a $Ga$-dependent threshold $Bo_o(Ga)$ decreasing from $Bo_o\approx0.3$ for $Ga=40$ to $Bo_o\approx0.1$ for $Ga=90$, an isolated bubble follows a zigzagging or spiralling path.  
Considering a fixed initial separation between the two bubbles and assuming their line of centres to be initially exactly vertical, the simulations allowed us to build a phase map gathering the encountered flow regimes in the $(Ga,Bo)$ plane. For Bond numbers less than the above threshold and $Ga\lesssim70$, i.e. weakly deformed bubbles and still moderate Reynolds numbers, the system evolves according to the ASE scenario discussed in detail for smaller $Ga$ in ZNM. Beyond a second threshold ranging from $Bo_c\approx0.5$ for $Ga\leq40$ to $Bo_c\approx0.7$ for $Ga\geq50$, the two bubbles always collide. We did not attempt to compute the collision process in detail, but analyzed the mechanisms that make the collision unavoidable under such conditions, despite the large lateral excursions of both bubbles. For Bond numbers in between the above two thresholds, the simulations revealed the existence of three markedly different regimes. The first of them arises for $Bo\gtrsim Bo_o$ and $Ga\lesssim50$. There, the initial lateral drift of the TB is succeeded by a `fully developed' stage in which the two bubbles rise independently within the same vertical plane along large-amplitude zigzagging paths, which defines the Coplanar Independent Zigzagging (CIZ) regime. Increasing the Bond number beyond the upper limit of this regime, one first encounters a Non-coplanar Independent Zigzagging (NIZ) regime. Here again, after the initial ASE stage, the two bubbles rise independently along zigzagging or flattened spiralling paths but the two paths stand in distinct vertical planes. The selection of the CIZ or NIZ regime depends on the ability of the LB to develop an unstable path on its own. If the aspect ratio of the LB is below $2.2$, its path can become unstable only due to the external disturbance provided by the lateral drift of the TB. This initial condition forces the two paths to take place within the same plane, leading to the CIZ regime. In contrast, when the oblateness of the LB is large enough, its path instability develops independently from the lateral motion of the TB. This makes the two bubbles rise in distinct planes, yielding the NIZ regime. Increasing the Bond number beyond the upper limit of that regime, but still for $Bo<Bo_c$, the system enters the Interacting Flattened Spiralling (IFS) regime. Here again, both bubbles follow zigzagging or flattened spiralling paths. However the attraction provided by the LB wake is strong enough for the TB to re-enter this wake one or more times in the course of its ascent. During such sequences, the trailing vortices shed by the LB directly hit the TB, which results in marked successive lateral deflections of its path. In the most inertial regimes considered here ($Ga\gtrsim70$), the CIZ regimes was not observed. Instead, for low Bond numbers ($Bo\lesssim0.1$, which at such $Ga$ is typically encountered in water and liquid metals), the bubble pair first follows an ASE scenario, beyond which the two paths perform small-amplitude erratic horizontal motions independently from each other. This defines the Small-amplitude Independent Erratic (SIE) regime, the existence of which results from the fact that the wake of an isolated bubble rising under such high-$Ga$ low-$Bo$ conditions is intrinsically stable but its path is not, the instability being then driven by the overall force and torque constraints imposed by Newton's second law. Increasing the Bond number beyond the upper limit of the SIE regime, the NIZ-IFS-collision sequence observed for $Ga=\mathcal{O}(50)$ beyond the CIZ regime is recovered.   \\
\indent We then showed that, even slight, a nonzero misalignment of the two bubbles from the reference in-line configuration favours oscillations within the same vertical plane. Then, depending on the lateral separation the two bubbles may reach, this constraint may promote or reduce the possibility for further interactions between them. Typically, these interactions weaken when the bubble oblateness is large ($Bo\approx0.5$), i.e. the system stands in the IFS regime for $\theta_0=0^\circ$. Indeed, the initial lateral drift of the TB takes place quite close to the LB in this case, so that the shear `felt' by the TB is strong and allows the minima reached by the lateral separation to be sufficiently large to avoid the TB to re-enter the LB wake. The reverse happens when the bubble oblateness is moderate ($Bo\approx0.3$), i.e. the system stands in the CIZ regime for $\theta_0=0^\circ$. There, as the initial drift of the TB takes place at quite large distances from the LB, the shear encountered by the TB during this stage is small and yields small minima of the lateral separation in later stages. Under such conditions, the TB returns repeatedly within the LB wake in the course of its oscillations and is directly hit by the trailing vortices every time the tandem is nearly vertical, maintaining the system in close interaction even in the long term. Intermediate scenarios may happen in the NIZ regime, with the geometrical constraint induced by the initial inclination promoting interactions between the two bubbles during some time, until a lateral deviation of the TB under the action of the LB wake makes the two planes of rise distinct and allows the two bubbles to continue their ascent independently. \\
\indent The role of the distance separating initially the two bubbles was found to depend dramatically on the considered regime. Variations in $\overline{S}_0$ only change the geometry of the final arrangement in the CIZ and NIZ regimes, i.e. for $Bo\lesssim0.4$, reducing (resp. increasing) the inclination of the bubble pair if the initial separation is increased (resp. reduced). Things are more complex in the IFS regime, i.e. for $Bo\approx0.5$ and $Ga\gtrsim50$, because the evolution of the system then crucially depends on both the intensity and the orientation of the interaction taking place between the TB and the LB wake when the two bubbles realign vertically for the first time. If this interaction is strong, the lateral deviation undergone by the TB is large enough for the two paths to remain sufficiently far apart that no direct interaction further happens. Increasing $\overline{S}_0$ weakens the above lateral deviation, thus maintaining the two paths closer to each other. This of course favours the possibility of new realignments, hence of further interactions. However these interactions weaken since the vertical separation between the two bubbles increases. So, while some increase beyond the reference value $\overline{S}_0=8$ tends to maintain permanently the system in the IFS regime, a larger increase makes it transition to the NIZ or CIZ regime after some time. For Bond numbers of $\mathcal{O}(1)$, the two bubbles collide no matter how large $\overline{S}_0$. With $\overline{S}_0\lesssim10$, the approach to collision is driven by the usual attraction mechanism resulting from the longitudinal pressure gradient along the LB wake. Although the lift force acting on the TB tends classically to extract it from that wake, this effect is too weak compared to the wake-induced suction to produce a sufficiently large deviation of the TB avoiding collision. For larger $\overline{S}_0$, the TB stands further downstream in the LB wake, so that the longitudinal pressure gradient (hence the attractive effect) is significantly weaker than in the previous case. In this regime, the Reynolds number and the bubble oblateness being large ($Re=\mathcal{O}(100),\,\chi\gtrsim2.5$) while the transverse shear perceived by the TB is small, the shear-induced lift force tends to push it toward the wake centreline, according to the reversal mechanism identified by \cite{adoua2009reversal}. As a consequence, the TB stays within the LB wake even for large $\overline{S}_0$. The residual longitudinal attraction is then sufficient to make collision inevitable again.  \\
\indent The results discussed in this paper reveal a rich and frequently nonintuitive phenomenology. Beyond the role of the control parameters $Ga,\,Bo,\,\theta_0$ and $\overline{S}_0$ already identified in ZNM for $Ga\leq30$, the large-amplitude lateral oscillations of zigzagging or spiralling bubbles rising at higher $Ga$ add an extra level of complexity. Retrospectively, one may consider that this is partly due to fact that the initial time and the growth rate of these oscillations differ generally for the two bubbles, owing to the difference in the initial disturbance that triggers the path instability of each of them. Because of these characteristics, the phase difference between the oscillations of the two bubbles may under certain circumstances bring the orientation of their line of centres back to the vertical or close to it at various stages of their ascent, allowing the trailing vortices present in the LB wake to hit the TB, thereby modifying its dynamics. 
\\
\indent Present results are of direct use to understand the behaviour of high-Reynolds-number bubbly plumes released from a single injection point, or from an array of injectors provided the distance between two of them is much larger than the bubble radius. In contrast, they represent only a first step toward a satisfactory understanding of the mechanisms at stake in nearly homogeneous high-Reynolds-number bubbly suspensions. Indeed, there is an infinity of possible initial orientations for each bubble pair in such flows, and the in-line configuration considered here is only one of them. In particular, bubble pairs released side by side may exhibit deeply different dynamics, as the `risk' of a head-on collision and the probability that the two bubbles follow zigzagging paths located within the same plane are much smaller in that case. Therefore, the side-by-side configuration requires specific investigations similar to the present one. 
Moreover, collisions are frequent in bubbly suspensions, as soon as the gas volume fraction exceeds a few percent. This makes it necessary to  determine the fate of colliding bubbles, {i.e.} whether they bounce or coalesce depending on the flow and initial conditions. This aspect was deliberately disregarded here for computational reasons but represents a major technical challenge, suggesting that the route toward a detailed understanding of the dynamics of high-Reynolds-number bubbly suspensions is still long. 
\section*{Acknowledgements.}
 \noindent The authors acknowledge the supports from the NSFC (Natural Science Foundation of China) under grants $11872296,\,51636009$ and $51806173$, and from the Fundamental Research Funds for the Central Universities (xhj032021011-02). J.Z. also acknowledges the Young Talent Support Plan of Xi'an Jiaotong University.
 \vspace{-5mm}
 \section*{Declaration of interests.}  
 \noindent The authors report no conflict of interest.
\appendix
\section{Grid convergence in a high-Reynolds-number case}\label{app:sec1}

Several tests were reported in ZNM to assess the quality of the computational results. In particular, a grid convergence study with minimum grid spacings $\Delta_{min} = R/34,\,R/68$ and $R/136$ showed that grid convergence is obtained with $\Delta_{min} = R/68$ at least up to $Ga = 30$, which corresponds to Reynolds numbers of $\mathcal{O}(100)$. However in the present investigation, low-$Bo$ bubble pairs with $Ga = 90$ reach Reynolds numbers of $\mathcal{O}(500)$. The corresponding boundary layer being typically twice as thin as in the previous case, an extra grid convergence test is in order. For this purpose, we select the parameters $Ga = 90,\,Bo = 0.05$, a configuration belonging to the SIE regime. In that case, the final Reynolds number of each bubble is about $470$, so that with $\Delta_{min} = R/68$ (resp. $R/136$), approximately three (resp. six) grid cells stand within the boundary layer. If grid convergence is reached, the rising speed of each bubble should be grid-independent, except during the early stage corresponding to the onset of the lateral drift of the TB across the LB wake. Indeed, the initial axial symmetry of the system being broken by an instability, the transition to a non-axisymmetric state is governed by the asymmetry of `natural' numerical disturbances present in the code. Since these disturbances depend on the grid resolution (among other things), the onset of the lateral motion varies with the spatial resolution and no grid independence may be expected during that stage. A direct consequence of this grid-dependent onset is that the vertical and horizontal separations between the two bubbles cannot be strictly grid-independent on the long term, as they result from the time integration of the bubble velocities (see Appendix A in ZNM for a detailed discussion).
\begin{figure}
\vspace{6mm}
  \centerline{\includegraphics[width=1.05\textwidth]{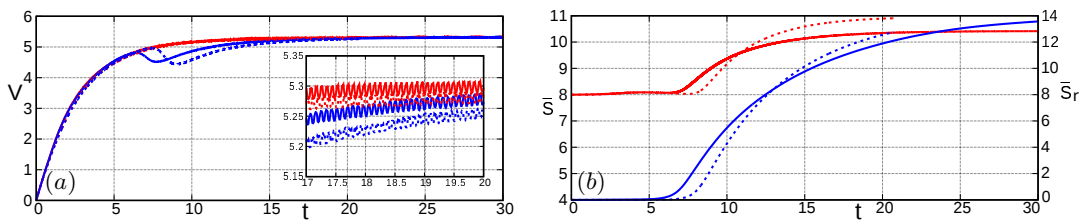}}
  \caption{Influence of grid resolution on the evolution of a bubble pair with $Ga = 90, Bo = 0.05$. $(a)$: Rising speed of the LB (red line) and TB (blue line); $(b)$: vertical (red line, left axis) and horizontal (blue line, right axis) components of the separation. Solid and dotted lines refer to simulations performed with $\Delta_{min} = R/68$ and $\Delta_{min} = R/136$, respectively. The inset in $(a)$ provides a precise view of the residual differences between the rising speeds predicted on the two grids at $t\approx20$. The small high-frequency oscillations visible in this inset result from the dynamic changes of the grid near the bubble surface in this low-$Bo$ high-$Ga$ regime.}
   \vspace{-37.5mm}
 \hspace{1.5mm}$(a)$ \hspace{65mm}$(b)$\\
   \vspace{30mm} 
\label{fa1}
\end{figure}
The test was carried out on the reference grid with $\Delta_{min} = R/68$ and on a twice as fine grid with $\Delta_{min} = R/136$. The computational time required on this refined grid is very large, owing to the larger number of cells and smaller time step. This is why the corresponding run was stopped at $t = 20$, beyond which the two bubbles rise independently. On both grids, the tolerance applied to the Poisson solver for the pressure field, defined as the maximum relative change of the fluid volume enclosed in a cell over one time step, was fixed to $1\times10^{-4}$. \\
\indent The results of this test are summarized in figure~\ref{fa1}. The inset in panel $(a)$ reveals that the rising speeds reached at $t=20$ on the two grids differ by less than $0.5\%$, and the difference is still reducing at that time. The bubble aspect ratio being close to $1.75$, the contribution of the boundary layer and wake to the drag, hence to the rising speed, is approximately $3.5\%$ at this Reynolds number \citep{Moore1965}. Therefore it can be concluded that this contribution, while small, is properly captured with $\Delta_{min} = R/68$, despite the limited number of cells standing within the boundary layer. This establishes the grid convergence of present high-$Ga$ results in the sense defined above. 
The differences observed on the rising speeds during the lateral drift of the TB and their consequences on the two components of the separation may be easily rationalized. The onset of the TB drift is seen to happen somewhat later on the finer grid. 
This delay implies that, on that grid, the vertical separation between the two bubbles is smaller when the TB starts drifting, resulting in a smaller vertical separation during some time (see the red curves in figure~\ref{fa1}$(b)$ in the time interval $7\lesssim t\lesssim10$). However this shorter $\overline{S}$ implies that the TB experiences a larger ambient shear during some time, hence a stronger lift force. This is why the growth rate of the horizontal separation is larger on the finer grid. 
Owing to this stronger transverse force, a larger fraction of the potential energy of the TB is converted into kinetic energy associated with the transverse motion, at the expense of the vertical motion. This is confirmed in figure~\ref{fa1}$(a)$, where the minimum rising speed of the TB during its lateral drift is seen to be slightly smaller on the finer grid. Due to this more severe transient reduction of the TB speed, the vertical separation observed on that grid becomes larger beyond $t\approx11$. The difference stops increasing at the end of the TB drift but persists in the long term, making the final $\overline{S}$ slightly larger on the finer grid.

\bibliographystyle{jfm}
\bibliography{bubble-inline}

\begin{thebibliography}{41}
\expandafter\ifx\csname natexlab\endcsname\relax\def\natexlab#1{#1}\fi
\def\au#1{#1} \def\ed#1{#1} \def\yr#1{#1}\def\at#1{#1}\def\jt#1{\textit{#1}}
  \def\bt#1{#1}\def\bvol#1{\textbf{#1}} \def\vol#1{#1} \def\pg#1{#1}
  \def\publ#1{#1}\def\arxiv#1{#1}\def\org#1{#1}\def\st#1{\textit{#1}}

\bibitem[Adoua {\em et~al.\/}(2009)Adoua, Legendre \&
  Magnaudet]{adoua2009reversal}
{\sc \au{Adoua, R.}, \au{Legendre, D.} \& \au{Magnaudet, J.}} \yr{2009}
  \at{Reversal of the lift force on an oblate bubble in a weakly viscous linear
  shear flow}.  \jt{J. Fluid Mech.}  \bvol{628},  \pg{23--41}.

\bibitem[Alm{\'e}ras {\em et~al.\/}(2015)Alm{\'e}ras, Risso, Roig, Cazin, Plais
  \& Augier]{Almeras2015}
{\sc \au{Alm{\'e}ras, E.}, \au{Risso, F.}, \au{Roig, V.}, \au{Cazin, S.},
  \au{Plais, C.} \& \au{Augier, F.}} \yr{2015}  \at{Mixing by bubble-induced
  turbulence}.  \jt{J. Fluid Mech.}  \bvol{776},  \pg{458--474}.

\bibitem[Auguste {\em et~al.\/}(2013)Auguste, Magnaudet \& Fabre]{Auguste2013}
{\sc \au{Auguste, F.}, \au{Magnaudet, J.} \& \au{Fabre, D.}} \yr{2013}
  \at{Falling styles of disks}.  \jt{J. Fluid Mech.}  \bvol{719},
  \pg{388--405}.

\bibitem[Auton(1987)]{auton1987lift}
{\sc \au{Auton, T.~R.}} \yr{1987}  \at{The lift force on a spherical body in a
  rotational flow}.  \jt{J. Fluid Mech.}  \bvol{183},  \pg{199--218}.

\bibitem[Blanco \& Magnaudet(1995)]{blanco1995structure}
{\sc \au{Blanco, A.} \& \au{Magnaudet, J.}} \yr{1995}  \at{The structure of the
  axisymmetric high-{R}eynolds number flow around an ellipsoidal bubble of
  fixed shape}.  \jt{Phys. Fluids}  \bvol{7},  \pg{1265--1274}.

\bibitem[Br{\"u}cker(1999)]{brucker1999structure}
{\sc \au{Br{\"u}cker, C.}} \yr{1999}  \at{Structure and dynamics of the wake of
  bubbles and its relevance for bubble interaction}.  \jt{Phys. Fluids}
  \bvol{11},  \pg{1781--1796}.

\bibitem[Bunner \& Tryggvason(2002)]{bunner2002dynamics}
{\sc \au{Bunner, B.} \& \au{Tryggvason, G.}} \yr{2002}  \at{Dynamics of
  homogeneous bubbly flows. {P}art 1. {R}ise velocity and microstructure of the
  bubbles}.  \jt{J. Fluid Mech.}  \bvol{466},  \pg{17--52}.

\bibitem[Bunner \& Tryggvason(2003)]{bunner2003effect}
{\sc \au{Bunner, B.} \& \au{Tryggvason, G.}} \yr{2003}  \at{Effect of bubble
  deformation on the properties of bubbly flows}.  \jt{J. Fluid Mech.}
  \bvol{495},  \pg{77--118}.

\bibitem[Cano-Lozano {\em et~al.\/}(2016)Cano-Lozano, Martinez-Bazan, Magnaudet
  \& Tchoufag]{cano2016paths}
{\sc \au{Cano-Lozano, J.~C.}, \au{Martinez-Bazan, C.}, \au{Magnaudet, J.} \&
  \au{Tchoufag, J.}} \yr{2016}  \at{Paths and wakes of deformable nearly
  spheroidal rising bubbles close to the transition to path instability}.
  \jt{Phys. Rev. Fluids}  \bvol{1},  \pg{053604}.

\bibitem[Chesters \& Hofman(1982)]{Chesters1982}
{\sc \au{Chesters, A.~K.} \& \au{Hofman, G}} \yr{1982}  \at{Bubble coalescence
  in pure liquids}.  \jt{Appl. Sci. Res.}  \bvol{38},  \pg{353--361}.

\bibitem[Duineveld(1998)]{Duineveld1998}
{\sc \au{Duineveld, P.~C.}} \yr{1998}  \at{Bouncing and coalescence of bubble
  pairs rising at high {R}eynolds number in pure water or aqueous surfactant
  solutions}.  \jt{Appl. Sci. Res.}  \bvol{58},  \pg{409--439}.

\bibitem[Esmaeeli \& Tryggvason(1999)]{Esmaeeli1999}
{\sc \au{Esmaeeli, A.} \& \au{Tryggvason, G.}} \yr{1999}  \at{Direct numerical
  simulations of bubbly flows {P}art 2. {M}oderate {R}eynolds number arrays}.
  \jt{J. Fluid Mech.}  \bvol{385},  \pg{325--358}.

\bibitem[Esmaeeli \& Tryggvason(2005)]{esmaeeli2005direct}
{\sc \au{Esmaeeli, A.} \& \au{Tryggvason, G.}} \yr{2005}  \at{A direct
  numerical simulation study of the buoyant rise of bubbles at {O}(100)
  {R}eynolds number}.  \jt{Phys. Fluids}  \bvol{17},  \pg{093303}.

\bibitem[Filella {\em et~al.\/}(2020)Filella, Ern \&
  Roig]{filella2020interaction}
{\sc \au{Filella, A.}, \au{Ern, P.} \& \au{Roig, V.}} \yr{2020}
  \at{Interaction of two oscillating bubbles rising in a thin-gap cell:
  vertical entrainment and interaction with vortices}.  \jt{J. Fluid Mech.}
  \bvol{888},  \pg{A13}.

\bibitem[Fortes {\em et~al.\/}(1987)Fortes, Joseph \&
  Lundgren]{fortes1987nonlinear}
{\sc \au{Fortes, A.~F.}, \au{Joseph, D.~D.} \& \au{Lundgren, T.~S.}} \yr{1987}
  \at{Nonlinear mechanics of fluidization of beds of spherical particles}.
  \jt{J. Fluid Mech.}  \bvol{177},  \pg{467--483}.

\bibitem[Gvozdic {\em et~al.\/}(2018)Gvozdic, Alm{\'e}ras, Mathai, Zhu, van
  Gils, Verzicco, Huisman, Sun \& Lohse]{Gvozdic2018}
{\sc \au{Gvozdic, B.}, \au{Alm{\'e}ras, E.}, \au{Mathai, V.}, \au{Zhu, X.},
  \au{van Gils, D. P.~M.}, \au{Verzicco, R.}, \au{Huisman, S.~G.}, \au{Sun, C.}
  \& \au{Lohse, D.}} \yr{2018}  \at{Experimental investigation of heat
  transport in homogeneous bubbly flow}.  \jt{J. Fluid Mech.}  \bvol{845},
  \pg{226--244}.

\bibitem[Hallez \& Legendre(2011)]{hallez2011interaction}
{\sc \au{Hallez, Y.} \& \au{Legendre, D.}} \yr{2011}  \at{Interaction between
  two spherical bubbles rising in a viscous liquid}.  \jt{J. Fluid Mech.}
  \bvol{673},  \pg{406--431}.

\bibitem[Innocenti {\em et~al.\/}(2021)Innocenti, Jaccod, Popinet \&
  Chibbaro]{innocenti2021direct}
{\sc \au{Innocenti, A.}, \au{Jaccod, A.}, \au{Popinet, S.} \& \au{Chibbaro,
  S.}} \yr{2021}  \at{Direct numerical simulation of bubble-induced
  turbulence}.  \jt{J. Fluid Mech.}  \bvol{918},  \pg{A23}.

\bibitem[Joseph {\em et~al.\/}(1986)Joseph, Fortes, Lundgren \&
  Singh]{joseph1986nonlinear}
{\sc \au{Joseph, D.~D.}, \au{Fortes, A.}, \au{Lundgren, T.~S.} \& \au{Singh,
  P.}} \yr{1986}  \at{Nonlinear mechanics of fluidization of spheres, cylinders
  and disks in water}.  \bt{In {\em Advances in Multiphase Flow and Related
  Problems\/} (ed. \ed{G.~Papanicolaou})},  \pg{pp. 101--122}.  \publ{SIAM}.

\bibitem[Kong {\em et~al.\/}(2019)Kong, Mirsandi, Buist, Peters, Baltussen \&
  Kuipers]{kong2019hydrodynamic}
{\sc \au{Kong, G.}, \au{Mirsandi, H.}, \au{Buist, K.~A.}, \au{Peters, E. A.
  J.~F.}, \au{Baltussen, M.~W.} \& \au{Kuipers, J. A.~M.}} \yr{2019}
  \at{Hydrodynamic interaction of bubbles rising side-by-side in viscous
  liquids}.  \jt{Exp. Fluids}  \bvol{60},  \pg{155}.

\bibitem[Kusuno \& Sanada(2021)]{kusuno2021wake}
{\sc \au{Kusuno, H.} \& \au{Sanada, T.}} \yr{2021}  \at{Wake-induced lateral
  migration of approaching bubbles}.  \jt{Int. J. Multiphase Flow}  \bvol{139},
   \pg{103639}.

\bibitem[Kusuno {\em et~al.\/}(2019)Kusuno, Yamamoto \& Sanada]{kusuno2019lift}
{\sc \au{Kusuno, H.}, \au{Yamamoto, H.} \& \au{Sanada, T.}} \yr{2019}  \at{Lift
  force acting on a pair of clean bubbles rising in-line}.  \jt{Phys. Fluids}
  \bvol{31},  \pg{072105}.

\bibitem[Loisy {\em et~al.\/}(2017)Loisy, Naso \& Spelt]{loisy2017buoyancy}
{\sc \au{Loisy, A.}, \au{Naso, A.} \& \au{Spelt, P. D.~M.}} \yr{2017}
  \at{Buoyancy-driven bubbly flows: ordered and free rise at small and
  intermediate volume fraction}.  \jt{J. Fluid Mech.}  \bvol{816},
  \pg{94--141}.

\bibitem[Magnaudet \& Mougin(2007)]{magnaudet2007wake}
{\sc \au{Magnaudet, J.} \& \au{Mougin, G.}} \yr{2007}  \at{Wake instability of
  a fixed spheroidal bubble}.  \jt{J. Fluid Mech.}  \bvol{572},  \pg{311--337}.

\bibitem[Moore(1965)]{Moore1965}
{\sc \au{Moore, D.~W.}} \yr{1965}  \at{The velocity of rise of distorted gas
  bubbles in a liquid of small viscosity}.  \jt{J. Fluid Mech.}  \bvol{23},
  \pg{749--766}.

\bibitem[Mougin \& Magnaudet(2002)]{mougin2001path}
{\sc \au{Mougin, G.} \& \au{Magnaudet, J.}} \yr{2002}  \at{Path instability of
  a rising bubble}.  \jt{Phys. Rev. Lett.}  \bvol{88},  \pg{014502}.

\bibitem[Mougin \& Magnaudet(2006)]{mougin2006wake}
{\sc \au{Mougin, G.} \& \au{Magnaudet, J.}} \yr{2006}  \at{Wake-induced forces
  and torques on a zigzagging/spiralling bubble}.  \jt{J. Fluid Mech.}
  \bvol{567},  \pg{185--194}.

\bibitem[Popinet(2009)]{popinet2009accurate}
{\sc \au{Popinet, S.}} \yr{2009}  \at{An accurate adaptive solver for
  surface-tension-driven interfacial flows}.  \jt{J. Comput. Phys.}
  \bvol{228},  \pg{5838--5866}.

\bibitem[Popinet(2015)]{popinet2015quadtree}
{\sc \au{Popinet, S.}} \yr{2015}  \at{A quadtree-adaptive multigrid solver for
  the {S}erre--{G}reen--{N}aghdi equations}.  \jt{J. Comput. Phys.}
  \bvol{302},  \pg{336--358}.

\bibitem[Riboux {\em et~al.\/}(2010)Riboux, Risso \& Legendre]{Riboux2010}
{\sc \au{Riboux, G.}, \au{Risso, F.} \& \au{Legendre, D.}} \yr{2010}
  \at{Experimental characterization of the agitation generated by bubbles
  rising at high {R}eynolds number}.  \jt{J. Fluid Mech.}  \bvol{643},
  \pg{509--539}.

\bibitem[Risso \& Ellingsen(2002)]{risso2002velocity}
{\sc \au{Risso, F.} \& \au{Ellingsen, K.}} \yr{2002}  \at{Velocity fluctuations
  in a homogeneous dilute dispersion of high-{R}eynolds-number rising bubbles}.
   \jt{J. Fluid Mech.}  \bvol{453},  \pg{395--410}.

\bibitem[Ryskin \& Leal(1984)]{Ryskin1984}
{\sc \au{Ryskin, G.} \& \au{Leal, L.~G.}} \yr{1984}  \at{Numerical solution of
  free-boundary problems in fluid mechanics. {P}art 2. {B}uoyancy-driven motion
  of a gas bubble through a quiescent liquid}.  \jt{J. Fluid Mech.}
  \bvol{148},  \pg{19--35}.

\bibitem[Sanada {\em et~al.\/}(2009)Sanada, Sato, Shirota \&
  Watanabe]{Sanada2009}
{\sc \au{Sanada, T.}, \au{Sato, A.}, \au{Shirota, M.} \& \au{Watanabe, M.}}
  \yr{2009}  \at{Motion and coalescence of a pair of bubbles rising side by
  side}.  \jt{Chem. Eng. Sci.}  \bvol{64},  \pg{2659--2671}.

\bibitem[Sangani \& Didwania(1993)]{Sangani1993}
{\sc \au{Sangani, A.~S.} \& \au{Didwania, A.~K.}} \yr{1993}  \at{Dynamic
  simulations of flows of bubbly liquids at large {R}eynolds numbers}.  \jt{J.
  Fluid Mech.}  \bvol{250},  \pg{307--337}.

\bibitem[Smereka(1993)]{Smereka1993}
{\sc \au{Smereka, P.}} \yr{1993}  \at{On the motion of bubbles in a periodic
  box}.  \jt{J. Fluid Mech.}  \bvol{254},  \pg{79--112}.

\bibitem[Stewart(1995)]{Stewart1995}
{\sc \au{Stewart, C.~W.}} \yr{1995}  \at{Bubble interaction in low-viscosity
  liquids}.  \jt{Int. J. Multiphase Flow}  \bvol{21},  \pg{1037--1046}.

\bibitem[Tchoufag {\em et~al.\/}(2014{\natexlab{{\em a\/}}})Tchoufag, Fabre \&
  Magnaudet]{Tchoufag2014b}
{\sc \au{Tchoufag, J.}, \au{Fabre, D.} \& \au{Magnaudet, J.}}
  \yr{2014{\natexlab{{\em a\/}}}}  \at{Global linear stability analysis of the
  wake and path of buoyancy-driven disks and thin cylinders}.  \jt{J. Fluid
  Mech.}  \bvol{740},  \pg{278--311}.

\bibitem[Tchoufag {\em et~al.\/}(2014{\natexlab{{\em b\/}}})Tchoufag, Magnaudet
  \& Fabre]{Tchoufag2014}
{\sc \au{Tchoufag, J.}, \au{Magnaudet, J.} \& \au{Fabre, D.}}
  \yr{2014{\natexlab{{\em b\/}}}}  \at{Linear instability of the path of a
  freely rising spheroidal bubble}.  \jt{J. Fluid Mech.}  \bvol{751},  \pg{R4}.

\bibitem[Yin \& Koch(2008)]{yin2008velocity}
{\sc \au{Yin, X.} \& \au{Koch, D.~L.}} \yr{2008}  \at{Velocity fluctuations and
  hydrodynamic diffusion in finite-{R}eynolds-number sedimenting suspensions}.
  \jt{Phys. Fluids}  \bvol{20},  \pg{043305}.

\bibitem[Zenit {\em et~al.\/}(2001)Zenit, Koch \& Sangani]{Zenit2001}
{\sc \au{Zenit, R.}, \au{Koch, D.~L.} \& \au{Sangani, A.~S.}} \yr{2001}
  \at{Measurements of the average properties of a suspension of bubbles rising
  in a vertical channel}.  \jt{J. Fluid Mech.}  \bvol{429},  \pg{307--342}.

\bibitem[Zenit \& Magnaudet(2008)]{Zenit2008}
{\sc \au{Zenit, R.} \& \au{Magnaudet, J.}} \yr{2008}  \at{Path instability of
  rising spheroidal air bubbles: a shape-controlled process}.  \jt{Phys.
  Fluids}  \bvol{20},  \pg{061702}.

\end{thebibliography}


\end{document}